\documentclass[twocolumn,showpacs,preprintnumbers,amsmath,amssymb,showkeys]{revtex4}

\usepackage{graphicx}
\usepackage{dcolumn}
\usepackage{bm}

\newcommand{\be}{\begin{eqnarray}}
\newcommand{\ee}{\end{eqnarray}}

\begin{document}

\preprint{JLAB-PHY-05-9}
\preprint{RM3-TH/04-24}

\title{Global Analysis of Data on the Proton Structure Function 
$\mathbf g_1$\\ and Extraction of its Moments\footnote{\bf To appear in Physical Review D.}\\}

\author{M.~Osipenko$^{1,2}$\footnote{e-mail: osipenko@ge.infn.it},
S.~Simula$^{3}$,
W.~Melnitchouk$^{4}$,
P.~Bosted$^{4}$,
V.~Burkert$^{4}$,
E.~Christy$^{5}$,
K.~Griffioen$^{6}$,
C.~Keppel$^{5}$,
S.~Kuhn$^{7}$,
G.~Ricco$^{1,8}$}

\address{\vspace{0.25cm}
         $^1$Istituto Nazionale di Fisica Nucleare, Sezione di Genova, 16146 Genova, Italy}
\address{$^2$Skobeltsyn Institute of Nuclear Physics, 119992 Moscow, Russia}
\address{$^3$Istituto Nazionale di Fisica Nucleare, Sezione Roma III, 00146 Roma, Italy}
\address{$^4$Thomas Jefferson National Accelerator Facility, Newport News, Virginia 23606}
\address{$^5$Hampton University, Hampton, Virginia, 23668}
\address{$^6$College of William and Mary, Williamsburg, Virginia 23187}
\address{$^7$Old Dominion University, Norfolk, Virginia 23529}
\address{$^8$Dipartimento di Fisica dell'Universit\`a, 16146 Genova, Italy}

\begin{abstract}
Inspired by recent measurements with the CLAS detector at Jefferson Lab,
we perform a self-consistent analysis of world data on the proton
structure function $g_1$ in the range $0.17 < Q^2 < 30$~(GeV/c)$^2$.
We compute for the first time low-order moments of $g_1$ and study their 
evolution from small to large values of $Q^2$. The analysis includes the 
latest data on both the unpolarized inclusive cross sections and the ratio 
$R = \sigma_L / \sigma_T$ from Jefferson Lab, as well as a new model for the 
transverse asymmetry $A_2$ in the resonance region. The contributions of 
both leading and higher twists are extracted, taking into account effects 
from radiative corrections beyond the next-to-leading order by means of 
soft-gluon resummation techniques. The leading twist is determined with 
remarkably good accuracy and is compared with the predictions obtained 
using various polarized parton distribution sets available in the 
literature. The contribution of higher twists to the $g_1$ moments is 
found to be significantly larger than in the case of the unpolarized 
structure function $F_2$.
\end{abstract}



\maketitle

\section{\label{sec:introduction} Introduction}

One of the fundamental characterizations of nucleon structure is the
distribution of the nucleon spin among its quark and gluon constituents.
The classic tool for studying the quark spin distributions
experimentally has been inclusive lepton scattering off polarized
protons and neutrons. These experiments have determined the $g_1$ structure 
function of the nucleon, which, in the framework of the na\"ive Quark-Parton 
Model (QPM), is proportional to the difference between the distributions of 
quarks with spins aligned and anti-aligned to the nucleon spin.
Surprisingly, one finds that only 20-30\% of the proton spin is carried
by quarks -- an observation which came to be known as the ``proton spin
crisis''. Considerable effort, both experimentally and theoretically, has
subsequently gone into understanding where the remaining fraction of
the proton spin resides -- see Ref.~\cite{REVIEW} for recent reviews.

In terms of kinematics, most of the experimental study has been
focused on the high-$Q^2$ region, where the QPM description is most
applicable, and in the region of intermediate and small Bjorken-$x$,
which is important for evaluating parton model sum rules such as the
Bjorken sum rule. Qualitatively new information on the proton spin 
structure can be obtained by studying the $g_1$ structure function in 
the region of large Bjorken-$x$, at moderate values of the squared 
four-momentum transfer $Q^2$, in the range from 1 to 5~(GeV/c)$^2$.
Such a kinematic region is characterized by the presence of nucleon
resonances which contribute to higher-twist effects in the structure 
functions.

According to the operator product expansion (OPE) in QCD, the 
$Q^2$-evolution of structure function moments can be described in terms 
of a $1 / Q^2$, or twist, expansion, where the leading twist [${\cal{O}}(1)$
in $1 / Q^2$] represents scattering from individual partons, while higher
twists [${\cal{O}}(1 / Q^2)$ and higher] appear due to correlations among
partons. The inclusion of the contribution from the nucleon resonance 
production regions is a relevant point of our study, because resonances and 
Deep Inelastic Scattering (DIS) are closely related by the phenomenon 
of local quark-hadron duality \cite{BloGil,Rujula,Ricco_dual}. The latter 
has been extensively investigated at Jefferson Lab (JLab) for the case of 
the unpolarized structure function $F_2$ of the proton \cite{hallc,f2mom-hc}. 
In the polarized case, the contribution of the $\Delta(1232)$ resonance 
makes the analysis rather more interesting: since this resonance gives rise 
to a negative contribution to the $g_1$ structure function, while $g_1$ 
at high $Q^2$ is positive, one expects a breaking of local duality to 
occur in the $\Delta$ region at least up to several (GeV/c)$^2$ 
\cite{simula_g1}.

In this paper we report the results of a self-consistent extraction of
the proton structure function $g_1(x,Q^2)$ and its moments from the
world data on the longitudinal polarization asymmetry $A_\parallel$.
The extraction is based on a unique set of inputs for the structure
function $F_2$, the ratio $R = \sigma_L/\sigma_T$ and the transverse
asymmetry $A_2$. The complete data set measured at Jefferson Lab
\cite{fatemi,osipenko,Hall-C-R}, which covers the entire resonance region
with high precision, allows for the first time the $Q^2$-evolution of the
$g_1$ moments to be accurately evaluated up to $n = 7$. The results for the 
first moment have been presented in Ref.~\cite{M1p}, where the twist-four 
matrix element was extracted, and the proton's color electric and magnetic 
polarizabilities determined. Here we give the details of our analysis for 
all the moments up to $n = 7$.

In Section~\ref{sec:Moments} we describe the OPE framework of the moments 
analysis for the polarized structure function $g_1$. In Section~\ref{sec:DataAnalysis} 
we discuss the extraction of $g_1$ from the longitudinal asymmetry $A_\parallel$. 
The evaluation of the moments of $g_1$ and their uncertainties is presented in 
Section~\ref{sec:DataAnalysis}, and the extraction of both leading and higher 
twists is described in Section~\ref{sec:Extraction}. Finally, conclusions from 
this study are summarized in Section~\ref{sec:Conclusions}.

\section{Moments of the Structure Function $g_1$}\label{sec:Moments}

The complete $Q^2$-evolution of the structure functions can be obtained
using the OPE \cite{OPE} of the time-ordered product of the two currents
which enter into the virtual photon--nucleon forward Compton scattering
amplitude,
 \be
    T[J(z) ~ J(0)] = \sum_{n, \alpha} ~ f_n^{\alpha}(-z^2) ~ z^{\mu_1}
    z^{\mu_2} ... z^{\mu_n} ~ O_{\mu_1 \mu_2 ... \mu_n}^{\alpha}
    \label{eq:current}
 \ee
where $O_{\mu_1 \mu_2 ... \mu_n}^{\alpha}$ are symmetric traceless
operators of dimension $d_n^{\alpha}$ and twist $\kappa_n^{\alpha}
\equiv d_n^{\alpha} - n$, with $\alpha$ labeling different operators
of spin $n$. In Eq.~(\ref{eq:current}), $f_n^{\alpha}(-z^2)$ are coefficient
functions, which are calculable in perturbative QCD (pQCD) at short
light-cone distances $z^2 = (ct)^2 - \vec{z}^2 \approx 0$. Since the 
imaginary part of the forward Compton scattering amplitude is
simply the hadronic tensor containing the structure functions measured
in DIS experiments, Eq.~(\ref{eq:current}) leads to the well-known twist 
expansion for the Cornwall-Norton (CN) moments of $g_1(x,Q^2)$~\cite{Corn-Nort,Wandzura},
 \be
    && M_n^{\rm CN}(Q^2) \equiv \int_0^1 dx ~ x^{n - 1} ~ g_1^N(x, Q^2)
\nonumber \\
    && = \sum_{\kappa = 2,4~\cdots}^{\infty} E_{n \kappa}[\mu,\alpha_s(Q^2)] 
    ~ O_{n \kappa}(\mu) ~ \left( {\mu^2 \over Q^2} \right)^{(\kappa - 2)/2} ~~~~
    \label{eq:CN}
 \ee
for $n = 1, 3, 5, \ldots$. Here $\mu$ is the renormalization scale, 
$O_{n \kappa}(\mu)$ are the (reduced) matrix elements of operators with 
definite spin $n$ and twist $\kappa$, containing information about the 
nonperturbative structure of the target, and $E_{n \kappa}(\mu, Q^2)$ are 
dimensionless coefficient functions, which can be expressed perturbatively 
as a power series of the running coupling constant $\alpha_s(Q^2)$. 

\indent In the Bjorken limit ($Q^2, \nu \to \infty$, with $x = Q^2/2M\nu$ 
fixed, where $\nu$ is the energy transfer and $M$ the nucleon mass), only
operators with spin $n$ contribute to the $n$-th CN moment (\ref{eq:CN}).
At finite $Q^2$, however, operators with different spins can contribute.
Consequently the $1 / Q^2$ expansion of the CN moment $M_n^{\rm CN}(Q^2)$ 
contain in addition target-mass terms, proportional to powers of $M^2 / 
Q^2$, which are formally leading twist and of pure kinematical origin.
It was shown by Nachtmann \cite{Nachtmann} in the unpolarized case,
and subsequently generalized to the polarized structure functions in
Ref.~\cite{Wandzura}, that even when $M^2 / Q^2$ is nonzero, the moments
can be redefined in such a way that only spin-$n$ operators contribute
to the $n$-th moment. This is achieved by defining the ``Nachtmann moments'' 
of $g_1$ as
 \be
    M_n(Q^2)
    &\equiv& \int_0^1 dx {\xi^{n+1} \over x^2} \nonumber\\
    &\times& \left\{ g_1(x, Q^2) \left[ {x \over \xi} - 
    {n^2 \over (n + 2)^2} {M^2 x^2 \over Q^2} {\xi \over x}
    \right] \right. \nonumber\\
    && -\left. g_2(x, Q^2) ~ {M^2 x^2 \over Q^2} {4n \over n 
    + 2} \right\} ~~~~
    \label{eq:i_nm1}
 \ee
where $\xi = 2x / \left( 1 + \sqrt{1 + 4 M^2 x^2 / Q^2} \right)$ is the
Nachtmann scaling variable. Note that the evaluation of the polarized moments 
$M_n(Q^2)$ requires the knowledge of both structure functions $g_1$ and $g_2$.
In the DIS regime the contribution of $g_2$ to Eq.~(\ref{eq:i_nm1}) turns out 
to be typically small (see Ref.~\cite{simula_g1}). On the other hand, in the 
nucleon resonance production region the impact of $g_2$ is expected to be more 
significant, and here the lack of experimental information on the structure 
function $g_2$ can lead to systematic uncertainties.

Since the moments in Eq.~(\ref{eq:i_nm1}) are totally inclusive, the
integral in the right hand side of Eq.~(\ref{eq:i_nm1}) contains also
the contribution from the elastic peak located at $x = 1$,
\be
   && g_1^{\rm el}(x,Q^2) = \delta (x-1) G_M(Q^2)\frac{G_E(Q^2)+\tau G_M(Q^2)}{2(1+\tau)}
   \label{eq:g1_el} \\
   && g_2^{\rm el}(x,Q^2) = \delta (x-1) \tau G_M(Q^2)\frac{G_E(Q^2)-G_M(Q^2)}{2(1+\tau)} ~~~~
   \label{eq:g2_el}
\ee
with $G_E$ $(G_M)$ the proton electric (magnetic) elastic form factor
and $\tau = Q^2 / 4 M^2$.

Note that the structure function moments include the resonance
production region at low $Q^2$ and high $x$, which would be otherwise
problematic to include in a twist analysis performed directly in
$x$-space. In addition, since target-mass corrections are by definition 
subtracted from the moments~(\ref{eq:i_nm1}), the twist expansion of the 
Nachtmann moments $M_n(Q^2)$ directly reveals information on the 
nonperturbative correlations between partons, without relying on 
specific assumptions about the $x$-shape of the leading twist.

\begin{figure}[htb]
\includegraphics[bb=2.5cm 12cm 20cm 17cm, scale=0.6]{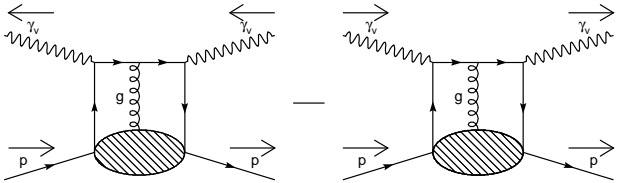}
\caption{\label{fig:TwistDiag} Example of a twist-4 diagram appearing in
the OPE of polarized structure function moments. Arrows indicate the spin 
projections of the particles.}
\end{figure}

For the leading twist contribution [$\kappa = 2$ in Eq.~(\ref{eq:CN})], one 
finds the well-known logarithmic $Q^2$ evolution of both singlet and non-singlet
moments. However, if one wants to extend the analysis to small $Q^2$ and 
large $x$, where the rest of the perturbative series becomes significant,
some procedure for the summation of higher orders of the pQCD expansion, 
such as  infrared renormalon models~\cite{renormalon,Ricco1} or soft-gluon 
resummation techniques~\cite{SGR,SIM00,alpha}, has to be applied. For higher 
twists, $\kappa >2$, the power-suppressed terms are related to quark-quark 
and quark-gluon correlations, as schematically illustrated in 
Fig.~\ref{fig:TwistDiag}, and should become important at small $Q^2$.

The evaluation of the Nachtmann moments (\ref{eq:i_nm1}) from available
data in the range $0.17 < Q^2 < 30$~(GeV/c)$^2$ will be described in
the next section. The OPE analysis of such experimental moments will allow 
us to extract simultaneously both the leading and the higher twist 
contributions. A precise evaluation would permit a comparison of the 
leading twist with the QCD predictions obtained from lattice simulations 
or with nonperturbative models of the nucleon.

\section{Data Analysis}\label{sec:DataAnalysis}

The data analysis was performed starting from measured longitudinal
proton asymmetries $A_\parallel$, which were converted into the
structure function $g_1$ using consistent values of the ratio
$R = \sigma_L/\sigma_T$ and the structure function $F_1$, as 
well as of the transverse proton asymmetry $A_2$. Our procedure 
is described in detail in the following. 

\subsection{\label{sec:intro}Asymmetry Database}

\begin{figure}[htb]
\begin{center}
\includegraphics[bb=0.25cm 5.5cm 20cm 23cm, scale=0.45]{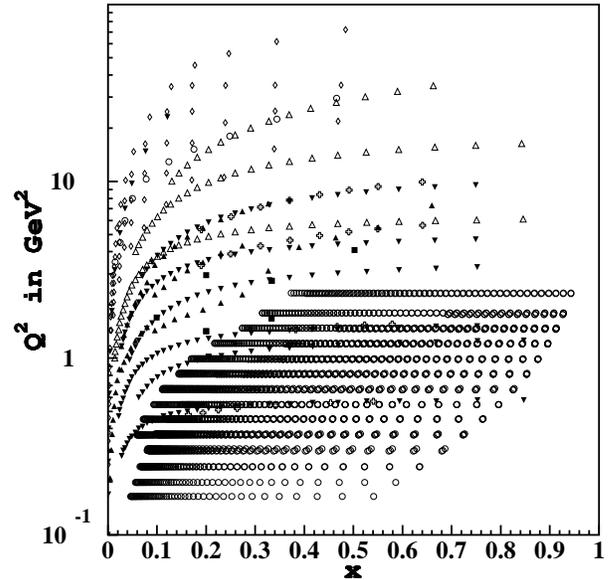}
\caption{\label{fig:kinem_al} Kinematics of $A_\parallel$ world data from
Refs.~\cite{fatemi,HERMES,SLAC-E155x,SLAC-E155,SLAC-E143,SLAC-E130,SLAC-E80,SMC-NA47,EMC-NA002}
(different symbols indicate different experiments).}
\end{center}
\end{figure}

\begin{figure}[htb]
\begin{center}
\includegraphics[bb=0.25cm 5.5cm 20cm 23cm, scale=0.45]{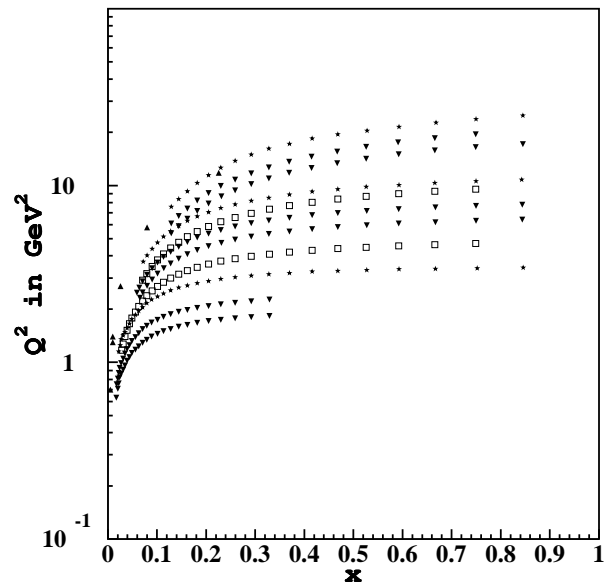}
\caption{\label{fig:atp_al} Kinematics of $A_\perp$ world data from
Refs.~\cite{HERMES},\cite{SLAC-E155x},\cite{SLAC-E155}b,\cite{SLAC-E80}b (different symbols 
indicate different experiments).}
\end{center}
\end{figure}

All available world data on the longitudinal and transverse asymmetries,
$A_\parallel$ and $A_\perp$, were collected from
Refs.~\cite{fatemi,HERMES,SLAC-E155x,SLAC-E155,SLAC-E143,SLAC-E130,SLAC-E80,SMC-NA47,EMC-NA002}
and Refs.~\cite{HERMES},\cite{SLAC-E155x},\cite{SLAC-E155}b,\cite{SLAC-E80}b, respectively. 
The full data set of $A_\parallel$ consists of two subsets corresponding to the resonance~\cite{fatemi} and DIS 
regions~\cite{HERMES,SLAC-E155x,SLAC-E155,SLAC-E143,SLAC-E130,SLAC-E80,SMC-NA47,EMC-NA002}.
The kinematic coverage of the experimental data is shown in
Figs.~\ref{fig:kinem_al} and \ref{fig:atp_al}, for $A_\parallel$ and
$A_\perp$, respectively. It can be seen that the resonance region is completely 
covered by the $A_\parallel$ data up to $Q^2 = 2.5$~(GeV/c)$^2$ with the 
inclusion of recent high quality data from CLAS~\cite{fatemi}.In contrast, 
the $A_\perp$ asymmetry is poorly determined in the resonance region.
The lack of data on $A_\perp$ here becomes problematic because of the
prominent role of the higher twist contributions at large values of $x$.

\subsection{\label{sec:extraction}Extraction of the Structure Function $g_1$}
In order to extract the structure function $g_1$ from the data collected
in our data base, one needs additional experimental inputs for the
structure function $F_1$, the ratio $R$, and the transverse asymmetry
$A_2$.  Indeed, the structure function $g_1$ is given by
 \be
    g_1(x, Q^2) & = & \frac{F_1(x, Q^2)}{1 + \gamma^2} \Biggl\{\frac{A_\parallel(x, Q^2)}{D}
    \nonumber \\
    &  & +\ (\gamma - \eta) A_2(x,Q^2) \Biggr\}\ ,
    \label{eq:g1ext}
 \ee
with
 \be
    \gamma = \frac{2Mx}{\sqrt{Q^2}} ,~~  
    \eta = \frac{\epsilon \sqrt{Q^2}}{E - \epsilon E^\prime} ,~~ 
    D = \frac{1 - \epsilon E^\prime / E}{1 + \epsilon R(x, Q^2)}\ ,
 \ee
where $E$ and $E^\prime$ are the incident and scattered electron energies
and $\epsilon$ is the virtual photon polarization. The ratio $R$ entering
above was taken from the parameterization given in Ref.~\cite{Hall-C-R}
for the resonance production region, while in the DIS domain the fit
R1998~\cite{r_fit_dis} was used.

Since the main goal of our analysis is a model independent extraction of
the moments of $g_1$, the structure function $F_1(x, Q^2)$ has been
obtained directly from experimental data. This has been possible because 
of the large amount of high quality data on the inclusive electron scattering 
cross section $d\sigma/d\Omega dE^\prime$ and on the structure function
$F_2$, covering both the resonance and DIS regions (for the list of data
used see Ref.~\cite{osipenko}). Therefore, for each point of the measured 
longitudinal asymmetry $A_\parallel$ we can find several nearby points 
with either $F_2$ or the inclusive cross section known from experiments.
For the interpolation of $F_1(x, Q^2)$ points, a simple procedure has been
used, which is described below. 

Having a data point with the measured $A_\parallel$ at some fixed $x_0$
and $Q^2_0$, we search in the combined database on the inclusive cross
section $d\sigma/d\Omega dE^\prime$ and the structure function $F_2$ for
several nearby experimental points. The search procedure chooses a 
rectangular bin around the point with coordinates ($x_0$, $Q^2_0$) of 
such a size that the selected area contains a number $N$ of experimental 
points either from $d\sigma / d\Omega dE^\prime$ or from $F_2$. The procedure 
then selects only those configurations whose number of points $N_{min} < N 
< N_{max}$, where $N_{min} = 2$ and $N_{max} = 6$ in the resonance region 
and $N_{min} = 1$ and $N_{max} = 4$ in the DIS case. Once a number of 
configurations have been collected (no more than 20 sets), the procedure 
looks for a minimum in the sum of the path integrals from each point 
$(x_i, Q^2_i)$ of measured $d\sigma / d\Omega dE^\prime$ or $F_2$ to the 
bin center ($x_0$, $Q^2_0$),
 \be
    \label{eq:path}
    S(x_0,Q^2_0)= \frac{1}{N F_1(x_0,Q^2_0)} \sum_i^N \int_{(x_i,Q^2_i)}^{(x_0,Q^2_0)} 
    dl\ |F_1(x,Q^2)|
 \ee
where the integral over $dl$ is taken along a straight line connecting 
the point $(x_i,Q^2_i)$ to the bin center $(x_0,Q^2_0)$. The structure 
function $F_1(x,Q^2)$ in this integral is constructed using the fits of 
$F_2$ from Ref.~\cite{f2_fit_dis} and of $R$ from Ref.~\cite{r_fit_dis} 
in DIS, while in the resonance production region $F_1$ is taken directly 
from Ref.~\cite{Hall-C-R}. The configuration selected is that which 
minimizes the function $S(x,Q^2)$ in Eq.~(\ref{eq:path}). 

From Fig.~\ref{fig:kinem_al}, and also from Fig.~1 of Ref.~\cite{osipenko}, 
one can see that in the resonance region, which is covered by the data from 
Ref.~\cite{fatemi}, the interpolation distances are very small, thanks to 
the measurements of inclusive cross section in the same kinematic 
range~\cite{osipenko,hallc}. A set of experimental points of $d\sigma / d\Omega 
dE^\prime$ or $F_2$ identified above is converted to the structure function 
$F_1$ according to
 \be
   F_1(x,Q^2) = \frac{M Q^2 E}{2\alpha^2 E^\prime} \frac{1-\epsilon}{1 + 
   \epsilon R(x,Q^2)} \frac{d\sigma}{d\Omega dE^\prime}
 \ee
and
 \be
    F_1(x,Q^2) = \frac{1 + 4 M^2 x^2 / Q^2}{2x (1 + R(x,Q^2))} F_2(x,Q^2) ~. 
 \ee
All the $F_1$ points obtained within the given bin are averaged together 
with their $x_i$ and $Q^2_i$ coordinates,
 \be
    && \overline{F_1}(x,Q^2) = \frac{1}{\delta^2} \sum_i \frac{F_1(x_i,Q^2_i)}{\delta_{F_1}^2(x_i,Q^2_i)}~, \\
    && \overline{x}   = \frac{1}{\delta^2} \sum_i \frac{x_i}{\delta_{F_1}^2(x_i,Q^2_i)}~, \\ 
    && \overline{Q^2} = \frac{1}{\delta^2} \sum_i \frac{Q^2_i}{\delta_{F_1}^2(x_i,Q^2_i)}~,
 \ee
where
 \be
    \delta = \sqrt{\sum_i  \frac{1}{\delta_{F_1}^2(x_i,Q^2_i)}}
 \ee
and $\delta_{F_1}$ is the statistical error of $F_1$. The mean value of 
$\overline{F_1}(x,Q^2)$ is then corrected by the bin centering correction 
using the models of Refs.~\cite{f2_fit_dis,r_fit_dis,Hall-C-R}. The value of 
the correction turns out to be very small with respect to statistical and 
systematic errors of the $A_\parallel$ data. Nevertheless, the correction 
value has been propagated in the total systematic error obtained for 
$\overline{F_1}$.

Once the transverse asymmetry $A_\perp$ is known, $A_2$ can be determined 
according to
 \be
    A_2 = \frac{1}{(1+\eta \zeta)} \Biggl[\frac{\zeta A_\parallel}{D} + 
    \frac{A_\perp}{d}\Biggr]\ , 
 \ee
where
 \be
    d = D\sqrt{\frac{2\epsilon}{1+\epsilon}}\ ,~~~~~ 
    \zeta = \eta \frac{1+\epsilon}{2\epsilon}\ .
 \ee
Since there are no experimental data on $A_\perp$ in the resonance region 
(see Fig.~\ref{fig:atp_al}), we consider several models:

\begin{itemize}

\item The model-independent constraint provided by the Soffer limit~\cite{Soffer}:
 \be
    |A_2| < \sqrt{\frac{A_1+1}{2}R} ~ .
 \ee
This inequality is exact and, provided $A_1$ and $R(x,Q^2)$ are measured, gives 
unambiguous limits.

\item Since it was shown in previous experiments that $A_2$ is in fact much 
smaller than the Soffer limit~\cite{SLAC-E155x}, one can simply assume $A_2 = 0$, 
with possible deviations from zero included in the systematic error.

\item In the present analysis we use a somewhat more sophisticated model 
for $A_2$ which is described in detail in Appendix~\ref{appendix1}.

\end{itemize}

\begin{figure}[htb]
\begin{center}
\includegraphics[bb=0cm 5cm 22cm 24cm, scale=0.45]{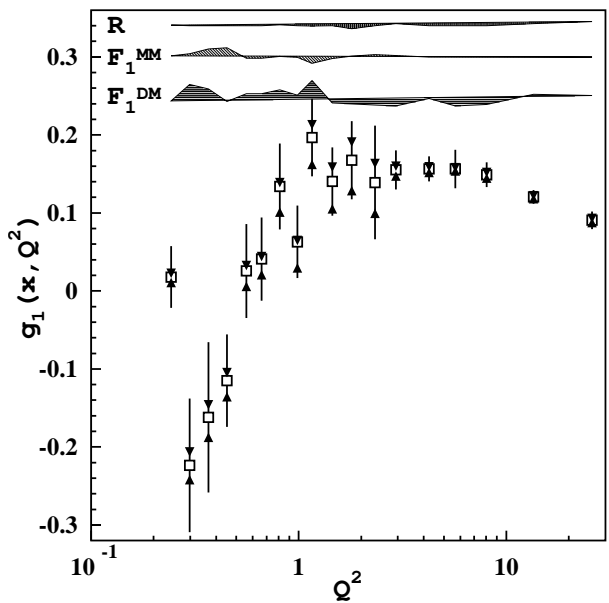}
\caption{\label{fig:g1_show} $Q^2$ dependence of the structure function $g_1$
at $x = 0.38 - 0.42$ obtained from the data in
Refs.~\cite{fatemi,HERMES,SLAC-E155x,SLAC-E155,SLAC-E143,SLAC-E130,SLAC-E80,SMC-NA47,EMC-NA002}
using the procedure described in the text. Open squares represent central values 
obtained with the $A_2$ model described in Appendix~\ref{appendix1}, while the 
filled triangles indicate upper and lower Soffer limits. The upper hatched area 
represents the difference between $g_1$ data points extracted with two different 
parameterizations of $R$ \cite{Hall-C-R,Ricco1}; middle hatched area $F_1^{MM}$ 
shows the difference between $g_1$ data points extracted using two different 
parameterizations of $F_1$ \cite{Hall-C-R,Bodek}; lower hatched area $F_1^{DM}$ 
shows the difference between $g_1$ extracted from the $F_1$ parameterization and 
data interpolation as described in the text.}
\end{center}
\end{figure}

The $Q^2$ dependence of $g_1(x,Q^2)$ at $x=0.38-0.42$ is shown in 
Fig.~\ref{fig:g1_show} using different assumptions about $A_2$ and $F_1$,
which provides an estimate of the systematic errors. The ranges and the 
averages for the various sources of systematic errors on $g_1$ are collected 
in Table~\ref{table:syserr}.

\begin{table}[htb]
\begin{center}
\caption{Range and average of systematic errors on $g_1$ (absolute value).}
\label{table:syserr}
\vspace{2mm}
\begin{tabular}{|c|c|c|} \hline
     Source of uncertainties         & Variation range      & Average \\ \hline
$A_\parallel$                        & 10$^{-4}$ -- 0.14~~   & 0.015  \\ \hline
$F_1$                                & 10$^{-7}$ -- 1.7~~~   & 0.014  \\ \hline
$\sigma_L / \sigma_T$                & 10$^{-4}$ -- 0.015    & 0.002  \\ \hline
$A_2$                                & 10$^{-7}$ -- 0.015    & 0.004  \\ \hline
Total                                & 10$^{-4}$ -- 1.7~~~   & 0.025  \\ \hline \end{tabular}
\end{center}
\end{table}

\subsection{Moments of the Structure Function $g_1$}\label{sec:d_nm}
As discussed in the introduction, the final goal of our data analysis
is the evaluation of the Nachtmann moments of the structure function $g_1$. 
The total Nachtmann moments were computed as the sum of the elastic 
($M_n^{\rm el}$) and inelastic ($M_n^{\rm in}$) moments,
 \be
    M_n(Q^2) = M^{\rm el}_n(Q^2) + M^{\rm in}_n(Q^2) ~ .
    \label{eq:d_nm_t1}
 \ee
The contribution from the elastic peak can be calculated by inserting
Eqs.~(\ref{eq:g1_el}, \ref{eq:g2_el}) into Eq.~(\ref{eq:i_nm1}),
 \be
    M^{\rm el}_n(Q^2) & = & \frac{\xi^n_{\rm el}}{2} G_M(Q^2) \Biggl\{
    \frac{G_E(Q^2) + \tau G_M(Q^2)}{1 + \tau} \nonumber \\
    && \times \Biggl[1-\frac{n^2}{(n+2)^2}\frac{M^2}{Q^2}\xi^2_{\rm el}\Biggl] 
    \nonumber \\
    & + & \frac{G_M(Q^2)-G_E(Q^2)}{1+\tau} \frac{n}{n+2}\xi_{\rm el} \Biggr\}~,
    \label{eq:d_em1}
 \ee
where $\xi_{\rm el} = 2 / (1 + \sqrt{1 + 1/ \tau})$.

The evaluation of the inelastic moments $M^{\rm in}_n$ involves the computation 
at fixed $Q^2$ of an integral over $x$. In practice the integral over $x$ was 
performed numerically using the standard trapezoidal method in the program 
TRAPER~\cite{cernlib}.

\begin{figure}
\includegraphics[bb=0cm 2cm 22cm 26cm, scale=0.4]{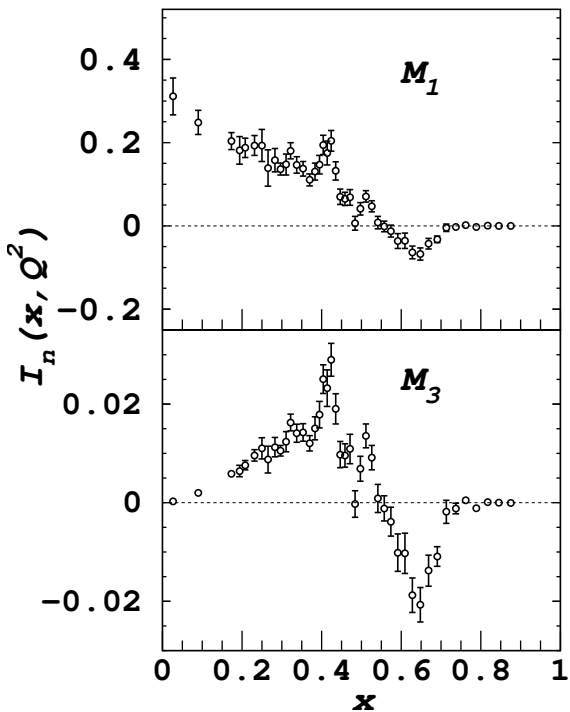}
\caption{\label{fig:intgr} Integrands of the Nachtmann moments at 
$Q^2 = 1$~GeV$^2$ for the $n = 1$ (upper) and the $n = 3$ (lower) moments.}
\end{figure}

The $Q^2$-range from 0.17 to 30 (GeV/c)$^2$ was divided into 24 bins 
increasing logarithmically with $Q^2$. Within each bin the world data were 
shifted to the central bin value $Q^2_0$ using the fit of $g_1^S(x, Q^2)$ 
from Ref.~\cite{simula_g1}, which covers both the resonance and DIS regions,
 \be
    g_1(x,Q^2_0) = g_1(x,Q^2)+\left(g_1^S(x,Q^2)-g_1^S(x,Q^2_0)\right) ~ .
    \label{eq:d_nm2}
 \ee
The difference between the actual and bin-centered data,
 \be
    \delta^{cent}_{g_1}(x,Q^2) = |g_1^S(x,Q^2_0)-g_1^S(x,Q^2)| ~ ,
 \ee
is added to the systematic error of $g_1$ in the Nachtmann moments 
extraction procedure. As an example, Fig.~\ref{fig:intgr} shows the 
integrands $I_n(x,Q^2)$ of two of the low-order moments as a function 
of $x$ at fixed $Q^2$. The significance of the large-$x$ region for 
higher moments can be clearly seen.

To obtain a data set dense in $x$, which reduces the error in the 
numerical integration, we performed an interpolation at each fixed 
$Q^2_0$ when two contiguous experimental data points differed by more 
than $\nabla$. The value of $\nabla$ depends on kinematics: in the 
resonance regions, where the structure function exhibits strong 
variations, $\nabla$ has to be smaller than half of the resonance 
widths, and is parameterized as $\nabla = 0.03 ~ M^2/Q^2$. Above the 
resonances, where $g_1$ is smooth, to account for the fact that
the available $x$ region decreases with decreasing $Q^2$, we set 
$\nabla = 0.1$. Finally, in the low $x$ region ($x<0.03$) where 
the $g_1$ shape depends weakly on $Q^2$, but strongly on $x$, we 
set $\nabla = 0.005$.

To fill the gap between two adjacent points $x_a$ and $x_b$, we used 
the interpolation function $g_1^{\rm int}(x,Q^2_0)$, defined as the 
parameterization from Ref.~\cite{simula_g1} offset to match the 
experimental data on both edges of the interpolating range. Assuming 
that the shape of the fit is correct, one has
 \be
    g_1^{\rm int}(x,Q^2_0)=\rho(Q^2_0)+g_1^S(x,Q^2_0)~,
    \label{eq:d_nm_i1}
 \ee
where the offset $\rho(Q^2_0)$ is defined as the weighted average, 
evaluated using all experimental points located within an interval 
$\Delta$ around $x_a$ or $x_b$:
 \be
    \rho(Q^2_0) & = & \delta_N^2(Q^2_0)\Biggl [\sum\limits_{i}^{|x_i-x_a| 
    < \Delta} \frac{g_1(x_i,Q^2_0)-g_1^S(x_i,Q^2_0)}{\bigl(\delta_{g_1}^{\rm stat}
    (x_i,Q^2_0)\bigr)^2} \nonumber \\
    & + & \sum\limits_{j}^{|x_j-x_b|<\Delta} \frac{g_1(x_j,Q^2_0) - 
    g_1^S(x_j,Q^2_0)}{\bigl(\delta_{g_1}^{\rm stat}(x_j,Q^2_0)\bigr)^2}\Biggr ]~,
    \label{eq:d_nm_i2}
 \ee
where $\delta_{g_1}^{\rm stat}(x_j,Q^2_0)$ is the $g_1$ statistical error and
 \be
    \delta_N(Q^2_0) & = & \Biggl [ \sum\limits_{i}^{|x_i-x_a|<\Delta} 
    \frac{1}{\bigl(\delta_{g_1}^{\rm stat}(x_i,Q^2_0)\bigr)^2} \nonumber \\
    & + & \sum\limits_{j}^{|x_j-x_b|<\Delta}\frac{1}
    {\bigl(\delta_{g_1}^{\rm stat}(x_j,Q^2_0)\bigr)^2} \Biggr ]^{-1/2} 
    \label{eq:d_nm_i3}
 \ee
is the statistical uncertainty of the normalization. Therefore, the statistical 
error of the moments calculated according to the trapezoidal rule~\cite{cernlib} 
was increased by adding the linearly correlated contribution from each interpolation 
interval as
 \be
    \delta^{\rm norm}_n (Q^2_0) & = & \delta_N (Q^2_0) \int\limits_{x_a}^{x_b} 
    dx \frac{\xi^{n+1}}{x^2} g_1^S(x,Q^2_0) \nonumber \\
    & \times & \Biggl[\frac{x}{\xi} - \frac{n^2}{(n+2)^2}
    \frac{M^2 x^2}{Q^2_0} \frac{\xi}{x} \Biggr] ~ .
    \label{eq:d_nm_i25}
 \ee

Since we average the difference $g_1(x_i,Q^2_0)-g_1^S(x_i,Q^2_0)$, $\Delta$ is 
not affected by the resonance structures, and its value is fixed to have more 
than two experimental points in most cases. Therefore, $\Delta$ is chosen to be 
equal to 0.15.

To fill the gap between the last experimental point and one of the 
integration limits ($x_a = 0$ or $x_b = 1$) we performed an extrapolation 
at each fixed $Q^2_0$ using $g_1^S(x,Q^2_0)$ including its uncertainty 
given in Ref.~\cite{simula_g1}. The results, together with their statistical 
and systematic errors, are presented in Table~\ref{table:r_nm1}.

\begin{table*}
\caption{\label{table:r_nm1}The inelastic Nachtmann moments for $n = 1, 3, 5$ and 
$7$ evaluated in the interval $0.17 \le Q^2 \le 30$~(GeV/c)$^2$. The moments were 
evaluated for $Q^2$ bins with more than 50\% data coverage. The data are reported 
together with the statistical and systematic errors; the low-$x$ extrapolation error 
is given for the first moment only (last number in the second column).}
\begin{ruledtabular}
\begin{tabular}{|c|c|c|c|c|} \cline{1-5}
$Q^2$~[(GeV/c)$^2$] & $M_1(Q^2) \times 10^{-3}$ & $M_3(Q^2) \times 10^{-4}$ & $M_5(Q^2) \times 10^{-5}$ & $M_7(Q^2) \times 10^{-6}$ \\ \cline{1-5}
 0.17 &--27.1 $\pm$  7   $\pm$ 12 $\pm$ 6   &--16.8 $\pm$ 2.5 $\pm$  5 & --8.5 $\pm$  1   $\pm$  2.5 &--4.8  $\pm$  0.6 $\pm$	1.3 \\ \cline{1-5}
 0.20 &--23.0 $\pm$  5   $\pm$  9 $\pm$ 6   &--17.0 $\pm$ 2   $\pm$  4 & --8.4 $\pm$  0.8 $\pm$  2   &--4.3  $\pm$  0.4 $\pm$	1.1 \\ \cline{1-5}
 0.24 & --4.2 $\pm$  4   $\pm$ 18 $\pm$ 7   &--16.1 $\pm$ 2   $\pm$ 11 &--11.0 $\pm$  1   $\pm$  7   &--7.3  $\pm$  0.7 $\pm$	4.5 \\ \cline{1-5}
 0.30 & --8.9 $\pm$  4   $\pm$ 19 $\pm$ 4   &--26.6 $\pm$ 2   $\pm$ 14 &--22.8 $\pm$  1.5 $\pm$ 11   &--18.8 $\pm$  1.2 $\pm$	9.3 \\ \cline{1-5}
 0.35 &   9.6 $\pm$  3   $\pm$ 12 $\pm$ 6   &--23.9 $\pm$ 2   $\pm$  8 &--28.9 $\pm$  2   $\pm$  7.5 &--31.2 $\pm$  1.5 $\pm$	7.4 \\ \cline{1-5}
 0.42 &  28.0 $\pm$  5   $\pm$ 11 $\pm$ 7   &--13.9 $\pm$ 4   $\pm$  9 &--26.6 $\pm$  4   $\pm$ 10   &--37.9 $\pm$  5	$\pm$  12   \\ \cline{1-5}
 0.50 &  36.3 $\pm$  4   $\pm$ 17 $\pm$ 3   &--13.2 $\pm$ 4   $\pm$ 16 &--31.0 $\pm$  5   $\pm$ 20   &--48.4 $\pm$  6	$\pm$  27   \\ \cline{1-5}
 0.60 &  43.4 $\pm$  3.5 $\pm$ 15 $\pm$ 4   &--12.2 $\pm$ 3   $\pm$ 16 &--35.9 $\pm$  4   $\pm$ 24   &--64.5 $\pm$  7	$\pm$  38   \\ \cline{1-5}
 0.70 &  56.0 $\pm$  3   $\pm$ 14 $\pm$ 6   &--0.1  $\pm$ 3   $\pm$ 18 &--28.4 $\pm$  4   $\pm$ 30   &--71.7 $\pm$  7	$\pm$  53   \\ \cline{1-5}
 0.84 &  69.0 $\pm$  3   $\pm$ 13 $\pm$ 1.5 &  15.3 $\pm$ 3   $\pm$ 19 &--8.7  $\pm$  5   $\pm$ 36   &--48.4 $\pm$ 11	$\pm$  74   \\ \cline{1-5}
 1.00 &  85.3 $\pm$  3   $\pm$ 11 $\pm$ 0.7 &  25.7 $\pm$ 2.5 $\pm$ 17 &--7.0  $\pm$  5   $\pm$ 37   &--81.1 $\pm$ 11	$\pm$  84   \\ \cline{1-5}
 1.20 &  94.2 $\pm$  3.5 $\pm$ 10 $\pm$ 1   &  53.7 $\pm$ 3   $\pm$ 17 &  57   $\pm$  7   $\pm$ 39   &  62.5 $\pm$ 18	$\pm$ 101   \\ \cline{1-5}
 1.40 & 102   $\pm$  4   $\pm$ 11 $\pm$ 2   &  68.6 $\pm$ 4   $\pm$ 20 &  88   $\pm$  7   $\pm$ 48   &  123  $\pm$ 19	$\pm$ 133   \\ \cline{1-5}
 1.70 & 114   $\pm$  3   $\pm$ 16 $\pm$ 2   &  92.9 $\pm$ 5   $\pm$ 20 &  150  $\pm$ 11   $\pm$ 48   &  295  $\pm$ 32	$\pm$ 142   \\ \cline{1-5}
 2.40 & 120   $\pm$  2.5 $\pm$  9 $\pm$ 3   &  108  $\pm$ 4   $\pm$ 16 &  218  $\pm$ 14   $\pm$ 46   &  572  $\pm$ 53	$\pm$ 152   \\ \cline{1-5}
 3.00 & 124   $\pm$  3   $\pm$  8 $\pm$ 3   &  107  $\pm$ 4   $\pm$ 10 &			     &  			    \\ \cline{1-5}
 3.50 & 113   $\pm$  7   $\pm$ 18 $\pm$ 1   &			       &			     &  			    \\ \cline{1-5}
 4.20 & 125   $\pm$  4   $\pm$  9 $\pm$ 3.5 & 110   $\pm$ 4.5 $\pm$  7 &			     &  			    \\ \cline{1-5}
 5.00 & 118   $\pm$  5   $\pm$ 11 $\pm$ 4   & 85.3  $\pm$ 7   $\pm$ 16 &  153  $\pm$ 18   $\pm$ 59   &  398  $\pm$ 61	$\pm$ 236   \\ \cline{1-5}
 6.00 & 122   $\pm$  5.5 $\pm$  8 $\pm$ 2   & 102   $\pm$ 6   $\pm$  8 &  219  $\pm$ 17   $\pm$ 18   &  664  $\pm$ 84	$\pm$  56   \\ \cline{1-5}
 8.40 &         			    & 102   $\pm$ 4   $\pm$  7 &			     &  			    \\ \cline{1-5}
10.00 & 128   $\pm$ 11   $\pm$ 13 $\pm$ 4   &			       &			     &  565  $\pm$ 85	$\pm$  66   \\ \cline{1-5}
15.50 & 130   $\pm$  3   $\pm$ 16 $\pm$ 4   & 88.8  $\pm$ 3   $\pm$ 16 &  187  $\pm$ 10   $\pm$ 30   &  597  $\pm$ 51	$\pm$  80   \\ \cline{1-5}
30.00 & 125   $\pm$  4   $\pm$ 10 $\pm$ 2.5 & 78.7  $\pm$ 5   $\pm$ 11 &  158  $\pm$ 20   $\pm$ 23   &  			    \\ \cline{1-5}
\end{tabular}
\end{ruledtabular}
\end{table*}

\subsection{Systematic Errors of the Moments}\label{sec:d_nm_ds}
The systematic error consists of experimental uncertainties in the data 
given in
Refs.~\cite{fatemi,HERMES,SLAC-E155x,SLAC-E155,SLAC-E143,SLAC-E130,SLAC-E80,SMC-NA47,EMC-NA002}
and uncertainties in the evaluation procedure. To estimate the first type 
of error we have to account for using many data sets measured at different 
laboratories and with different detectors. In the present analysis we assume 
that different experiments are independent and therefore only systematic errors 
within a particulardata set are correlated.

An upper limit for the contribution of the systematic error from each data set 
was thus evaluated as follows:

\begin{itemize}

\item we first applied a simultaneous shift to all experimental points 
in the data set by an amount equal to their systematic error;

\item the inelastic $n$-th moment obtained using these distorted 
data $\widetilde{M}^{\rm in}_{n(i)}(Q^2)$ is then compared to the 
original moments $M^{\rm in}_n(Q^2)$ evaluated with no systematic shifts;

\item finally, the deviations for each data set were summed in quadrature 
as independent values,
 \be
    \delta_n^D(Q^2) = \sqrt{\sum\limits_i^{N_{S}} 
    \Bigl(\widetilde{M}^{\rm in}_{n(i)}(Q^2)-M^{\rm in}_n(Q^2) \Bigr)^2}~,
    \label{eq:d_nm_ds1}
 \ee
where $N_{S}$ is the number of available data sets. The resulting error 
is summed in quadrature with $\delta_n^{\rm norm}(Q^2)$ to get the total 
systematic error on the \mbox{$n$-th} moment.

\end{itemize}

The second type of error is related to the bin centering, interpolation and 
extrapolation. The bin centering systematic uncertainty was estimated as
 \be
    \delta^C_n(Q^2) = \sum_i K_n(x_i,Q^2) w_i(Q^2) \delta^{cent}_{g_1}(x_i,Q^2)~,
 \ee
where, according to the Nachtmann moment definition and the trapezoidal 
integration rule, one has
 \be
    K_n(x_i,Q^2) & = &\frac{\xi_i^{n+1}}{x_i^2} g_1(x,Q^2)
    \Biggl[\frac{x_i}{\xi_i} - \frac{n^2}{(n+2)^2} \frac{M^2 x_i^2}{Q^2}
    \frac{\xi_i}{x_i}\Biggr]\ , \nonumber \\
    w_i(Q^2) &=& (x_{i+1}-x_{i-1})/2~.
 \ee

The systematic error of the interpolation was estimated by considering
the possible change of the fitting function slope in the interpolation 
interval, and was evaluated as a difference in the normalization at 
different edges:
 \be
    \delta_S(Q^2_0) & = & \Biggl | \frac{1}{N_i}\sum\limits_{i}^{|x_i-x_a|<\Delta}
    (g_1(x_i,Q^2_0)-g_1^S(x_i,Q^2_0)) \nonumber \\
    & - & \frac{1}{N_j}\sum\limits_{j}^{|x_j-x_b|<\Delta}
    (g_1(x_j,Q^2_0)-g_1^S(x_j,Q^2_0)) \Biggr | , ~~~~
    \label{eq:d_nm_i4}
 \ee
where $N_i$ and $N_j$ are the number of points used to evaluate the sums.
Since the structure function $g_1(x, Q^2)$ is a smooth function of $x$ 
below resonances, on the limited $x$-interval (smaller than $\nabla$) the 
linear approximation gives a good estimate. Thus, the error given in 
Eq.~(\ref{eq:d_nm_i4}) accounts for such a linear mismatch between the 
fitting function and the data on the interpolation interval. Meanwhile, 
the CLAS data cover all the resonance region and no interpolation was 
used there. The total systematic error introduced in the corresponding 
moment by the interpolation can therefore be estimated as
 \be
    \delta^I_n (Q^2_0) & = & \delta_S (Q^2_0) \int\limits_{x_a}^{x_b} dx \frac{\xi^{n+1}}{x^2} 
    g_1^S(x,Q^2) \nonumber \\  
    & \times & \Biggl[\frac{x}{\xi} - \frac{n^2}{(n+2)^2} \frac{M^2 x^2}{Q^2} 
    \frac{\xi}{x} \Biggr] ~ . \label{eq:d_nm_i5}
 \ee
The systematic errors obtained by these procedures are then summed in 
quadrature to give,
 \be
    \delta^P_n(Q^2) = \sqrt{(\delta^D_n(Q^2))^2+(\delta^C_n(Q^2))^2+(\delta^I_n(Q^2))^2}~.
    \label{eq:d_nm_i55}
 \ee

\begin{figure}
\includegraphics[bb=0cm 6cm 20cm 23cm, scale=0.45]{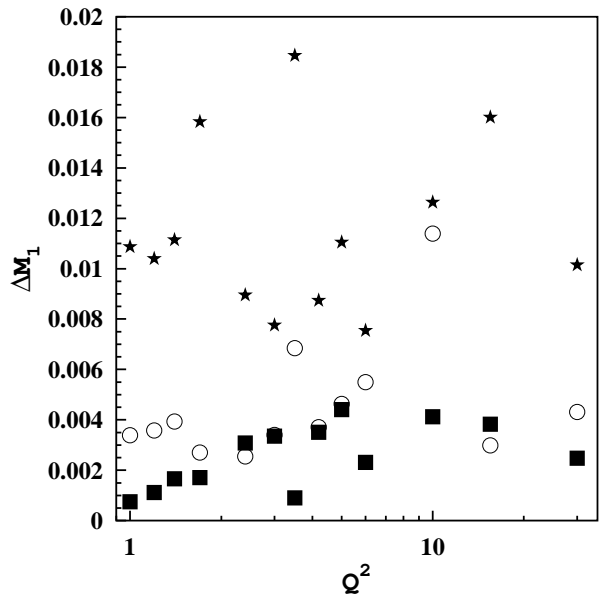}
\caption{\label{fig:SepErr} Errors of the inelastic Nachtmann moment 
$M_1$: the empty circles represent statistical errors; the stars show the 
systematic error obtained in Eq.~(\ref{eq:d_nm_i55}); the low-x extrapolation 
error is indicated by filled squares.}
\end{figure}

In order to study the systematic error on the extrapolation at very low 
$x$ we compared the moments extracted using different parameterizations 
of $g_1$. We choose a Regge inspired form from Ref.~\cite{simula_g1} and 
two QCD fits from Refs.~\cite{GRSV,DS}. The difference was significant only 
for $M_1$, for which the various errors are shown in Fig.~\ref{fig:SepErr} 
and separately given in Table~\ref{table:r_nm1}.

\begin{figure*}[htb]

\includegraphics[bb=1cm 10cm 20cm 29cm, scale=0.75]{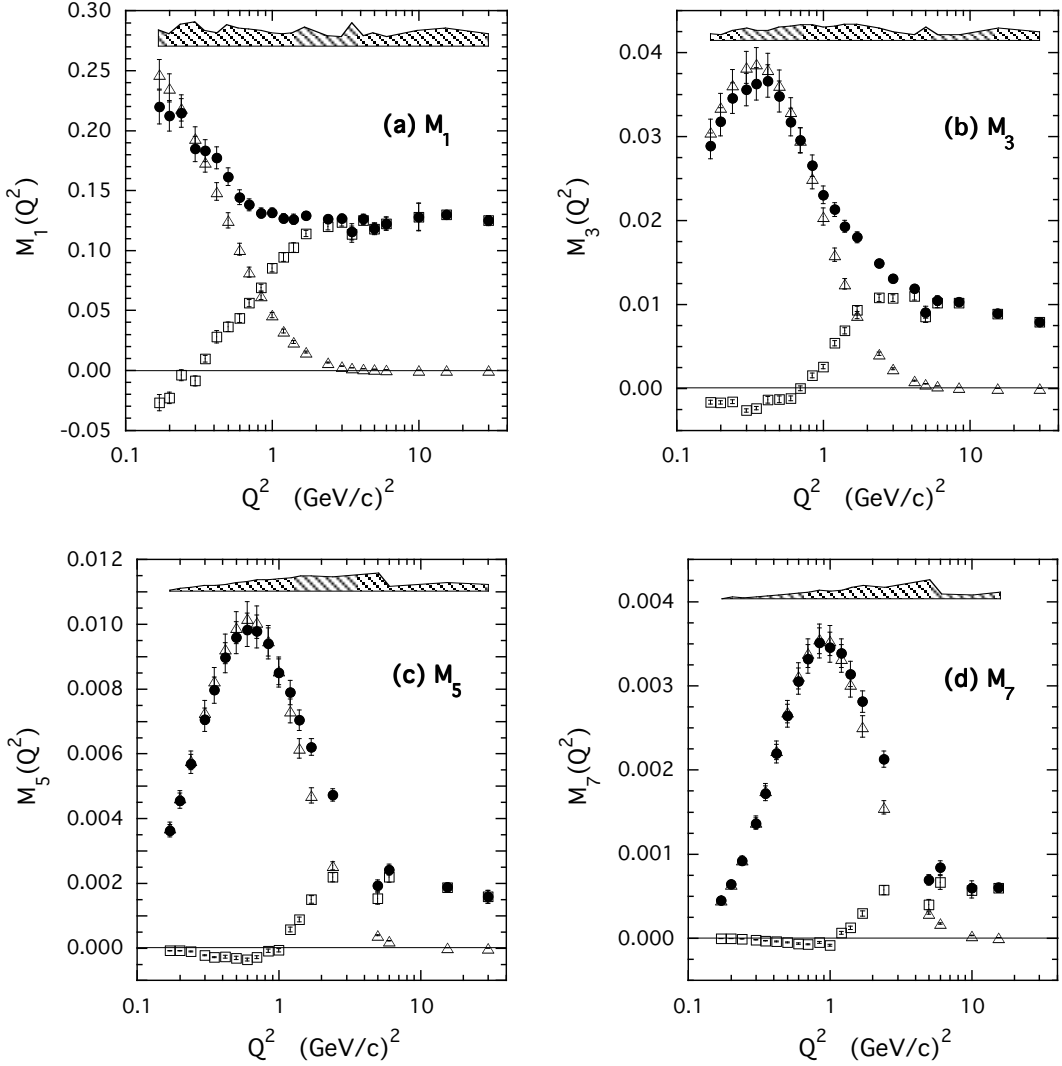}

\caption{\label{fig:M1n_dati} Total (inelastic + elastic) Nachtmann moments
$M_n(Q^2)$ (filled circles) [see Eq.~(\ref{eq:d_nm_t1})] extracted from 
the proton world data in the range $0.17 \leq Q^2 \leq 30$~(GeV/c)$^2$ for
$n = 1, 3, 5$ and $7$. Open squares and triangles correspond to the 
inelastic and elastic contributions, respectively. Statistical errors are 
reported for all the three terms; in case of the total moments the 
systematic errors are represented by the shaded bands.}

\end{figure*}

\indent According to Eq.~(\ref{eq:d_nm_t1}) the contribution from the proton
elastic peak should be added to the inelastic moments obtained above. The 
$Q^2$-dependence of the proton elastic form factors is parameterized as in 
Ref.~\cite{Bosted}, modified accordingly to the recent data on $G_E / 
G_M$~\cite{hall-a}, as described in Ref.~\cite{CLAS_note}. The uncertainty 
on the form factors is taken to be equal to 3\% according to the analysis of 
Ref.~\cite{Bosted}, and is added quadratically to both the statistic and the 
systematic errors. The elastic contribution $M_n^{\rm el}(Q^2)$ turns out to 
be a quite small correction for $Q^2 \gtrsim n ~ ({\rm GeV/c})^2$. Our final 
results for the total (inelastic + elastic) moments with $n = 1, 3, 5$ and $7$ 
are shown in Fig.~\ref{fig:M1n_dati}. Note also that the amount of the measured 
experimental contribution to $M_n(Q^2)$ is at least 50\%, and the systematic 
uncertainties increase significantly as $Q^2$ increases. 

\section{Extraction of leading and higher twists}\label{sec:Extraction}

In this section we present our analysis of the moments $M_n(Q^2)$ 
with $n > 1$. We extract both the leading and higher twist contributions 
to the moments, including a determination of the effective anomalous 
dimensions.

Results for the first moment $M_1(Q^2)$ were presented in Ref.~\cite{M1p}.
There the highest $Q^2$-points [$Q^2 > 5 ~ ({\rm GeV/c})^2$] were used to 
obtain the singlet axial charge, which for the renormalization group 
invariant definition in the $\overline{{\rm MS}}$ scheme (which is 
adopted throughout this paper) gave: $a_0^{\rm inv} = 0.145 \pm 
0.018(\mbox{{\rm stat.}}) \pm 0.103(\mbox{{\rm sys.}}) \pm 0.041({\rm low} x) 
\pm_{0.010}^{0.006}(\alpha_s)$, where the first and second errors are 
statistical and systematic, the  third is from the $x \to 0$ extrapolation, 
and the last is due to the uncertainty in $\alpha_s$.
From the $Q^2$-dependence of the first moment the matrix elements of 
twist-4 operators were extracted, which allowed a precise determination 
of the color electric and magnetic polarizabilities of the proton (see 
Ref.~\cite{M1p} for details).

As has been discussed in Refs.~\cite{Ricco1,SIM00,simula_g1,osipenko},
the extraction of higher twists at large $x$ is sensitive to the effects 
of high-order pQCD corrections, for both the polarized and unpolarized
cases. In particular, the use of the next-to-leading order (NLO) approximation 
for the leading twist is known to lead to unreliable results for the 
determination of the higher twists in the proton $F_2$ at large $x$ 
\cite{SIM00}. In this work we follow Refs.~\cite{SIM00,simula_g1,osipenko}, 
where the pQCD corrections beyond the NLO are estimated according to soft 
gluon resummation (SGR) techniques \cite{SGR} and a pure non-singlet (NS) 
evolution is assumed for $n \geq 3$~\footnote{This approximation is reasonable 
because of the effective decoupling of the pQCD evolution of the singlet 
quark and gluon densities at large $x$.}. However, in contrast to 
Refs.~\cite{SIM00,simula_g1,osipenko}, where SGR was considered for the 
quark coefficient function only, we consistently add in this work the 
resummation of large-$n$ logarithms appearing also in the one-loop and 
two-loop NS anomalous dimensions. This was previously used in Ref.~\cite{alpha} 
to determine the strong coupling constant $\alpha_s(M_Z^2)$ from the experimental 
moments of the proton $F_2$ structure function determined in Ref.~\cite{osipenko}.

Within the above framework, the Nachtmann moment of the leading twist part 
of the $g_1$ structure function, $\delta \eta_n(Q^2)$, is (for $n \geq 3$) 
explicitly given by
 \be
    \delta \eta_n(Q^2) & = & \delta A_n \left[ \alpha_s(Q^2) \right]^{\gamma_n^{{\rm NS}}} 
    \left\{ {\alpha_s(Q^2) \over 4 \pi} \delta R_n^{\rm NS} + e^{G_n(Q^2)} \right. 
    \nonumber \\
    & \times & \left. \left[ 1 + {\alpha_s(Q^2) \over 4 \pi} \left( 2 
    C_{\rm DIS}^{({\rm NLO})} + \Delta \gamma_{\rm DIS}^{(1, {\rm NS})} \right) \right] 
    \right\}  ~~~~
    \label{eq:SGR}
 \ee
where the constant $\delta A_n$ is defined to be the $n$-th moment of the 
leading twist at the renormalization scale $\mu^2$, and $\gamma_n^{\rm NS}$ 
is the one-loop NS anomalous dimension. In Eq.~(\ref{eq:SGR}) the quantity 
$\delta R_n^{\rm NS}$ is given by
 \be
    \delta R_n^{\rm NS} & = & 2 \left[ \delta C_n^{({\rm NLO})} - C_{{\rm DIS}}^{({\rm NLO})} - 
    C_{n, {\rm LOG}}^{({\rm NLO})} \right] \nonumber \\
    & + & \Delta \gamma_n^{(1, {\rm NS})} - \Delta \gamma_{{\rm DIS}}^{(1, {\rm NS})} - \Delta \gamma_{n, {\rm LOG}}^{(1, {\rm NS})}
    \label{eq:RNS}
 \ee
where 
 \be
     \Delta \gamma_n^{(1, {\rm NS})} \equiv \gamma_n^{(1, {\rm NS})}
     - {\beta_1 \over \beta_0} \gamma_n^{\rm NS}
     \label{eq:deltagamma}
 \ee
with $\gamma_n^{(1, {\rm NS})}$ being the two-loop NS anomalous 
dimension, $\beta_0 = 11 - 2 N_f / 3$, $\beta_1 = 102 - 38 N_f / 3$ and 
$N_f$ the number of active quark flavors at the scale $Q^2$.

\indent In Eq.~({\ref{eq:RNS}) $\delta C_n^{({\rm NLO})}$ is the NLO part
of the quark coefficient function, which in the $\overline{{\rm MS}}$ 
scheme is given by
 \be
    \delta C_n^{({\rm NLO})} &=& C_F \left\{ S_1(n) \left[ S_1(n) + {3 \over 2} - 
    {1 \over n(n+1)} \right] \right. \nonumber \\
    & & \left. - ~ S_2(n) + {1 \over 2n} + {1 \over n+1} + {1 \over n^2} - 
    {9 \over 2} \right\}
    \label{eq:Cn}
 \ee
where $C_F = (N_c^2 - 1) / (2 N_c)$ and $S_k(n) = \sum_{j=1}^n 1 /j^k$.
For large $n$ (corresponding to the large-$x$ region) the coefficient 
$C_n^{({\rm NLO})}$ is logarithmically divergent; indeed, since
$S_1(n) = \gamma_E + \mbox{log}(n) + {\cal O}(1/n)$,
where $\gamma_E = 0.577216$ is the Euler-Mascheroni constant, and 
$S_2(n) = \pi^2/6 + {\cal O}(1/n)$, one gets
 \be
    \delta C_n^{({\rm NLO})} & = & C_{{\rm DIS}}^{({\rm NLO})}
    + C_{n, {\rm LOG}}^{({\rm NLO})} + {\cal O}(1/n) ~ ,
 \ee
with
 \be
    C_{\rm DIS}^{({\rm NLO})} & = & C_F \left[ \gamma_E^2 + {3 \over 2} \gamma_E - 
    {9 \over 2} - {\pi^2 \over 6} \right] ~ ,
 \ee
and
 \be
    C_{n, {\rm LOG}}^{({\rm NLO})} & = & C_F ~ \mbox{ln}(n) \left[ \mbox{ln}(n) + 
    2\gamma_E + {3 \over 2} \right] ~ .
    \label{eq:Cn_exp}
 \ee
For the quantity $\Delta \gamma_n^{(1, {\rm NS})}$ in Eq.~(\ref{eq:deltagamma}) 
one obtains
 \be
    \Delta \gamma_n^{(1, {\rm NS})} & = & \Delta \gamma_{\rm DIS}^{(1, {\rm NS})}
    + \Delta \gamma_{n, {\rm LOG}}^{(1, {\rm NS})} + {\cal O}(1 / n) ~ ,
    \label{eq:gammas}
 \ee
where
 \be
    \Delta \gamma_{\rm DIS}^{(1, {\rm NS})} & = & {C_F \over \beta_0} 
    \left\{ C_F \left[ 2\pi^2 + 32 \tilde{S}(\infty) - 4 S_3(\infty) - 
    {3 \over 2} \right] \right. \nonumber \\
    & + & \left. C_A \left[ -{22 \over 9} \pi^2 - 16 \tilde{S}(\infty) - 
    {17 \over 6} \right] \right. \nonumber \\
    & + & \left. N_f \left[ {4 \pi^2 \over 9} + {1 \over 3} \right] 
    + \gamma_E \left( 8 K - 4 {\beta_1 \over \beta_0} \right) \right. \nonumber \\
    & + & \left. 3 {\beta_1 \over \beta_0} \right\} ~ ,
 \ee
and
 \be
    \Delta \gamma_{n, {\rm LOG}}^{(1, {\rm NS})} & = & {C_F \over \beta_0} 
    \left[ 8 K - 4 {\beta_1 \over \beta_0} \right] \mbox{ln}(n) ~ ,
    \nonumber \\
    &&
 \ee
with $C_A = N_c$, $\tilde{S}(\infty) = \sum_{j = 1}^{\infty} (-1)^j S_1(j) 
/ j^2 = -0.751286$, $S_3(\infty) = 1.202057$ and $K = C_A ~ (67/18 - \pi^2 
/ 6) - 5 N_f / 9$.

\indent In Eq.~(\ref{eq:SGR}) the function $G_n(Q^2)$ is the key 
quantity of the soft gluon resummation. At next-to-leading log (NLL)
accuracy one has 
 \be
    G_n(Q^2) = \mbox{ln}(n) ~ G_1(\lambda_n) + G_2(\lambda_n) + O\left[ 
    \alpha_s^k \mbox{ln}^{k-1}(n) \right] , ~~
    \label{eq:Gn}
 \ee
where $\lambda_n \equiv \beta_0 ~ \alpha_s(Q^2) ~ \mbox{ln}(n) / 4\pi$ and
 \be
    G_1(\lambda) & = & C_F {4 \over \beta_0 \lambda} \left[ \lambda + (1 
    -  \lambda) \mbox{ln}(1 - \lambda) \right] ~ , \nonumber \\[3mm]
    G_2(\lambda) & = & - C_F {4 \gamma_E + 3 \over \beta_0} \mbox{ln}(1 
    - \lambda) - C_F {8 K \over \beta_0^2} \mbox{ln}(1 - \lambda) 
    \nonumber \\
    & + & C_F {4  \beta_1 \over \beta_0^3} \mbox{ln}(1 - \lambda) \left[ 1 + 
    {1 \over 2} \mbox{ln}(1 - \lambda) \right] ~ .
    \label{eq:G1G2}
 \ee
Note that the function $G_2(\lambda)$ is divergent for $\lambda \to 1$; this means 
that at large $n$ ({\em i.e.},~large $x$) SGR cannot be extended to arbitrarily low 
values of $Q^2$. Therefore, to be sure that the SGR technique can be used reliably 
at NLL accuracy it is essential to check that $\lambda_n$ is small enough, 
which in our case means restricting the twist analysis to the $Q^2$-range 
above $0.8 \div 1$~(GeV/c)$^2$.

It is straightforward to see that in the limit $\lambda_n \ll 1$ one has 
$G_n(Q^2) \to \alpha_s(Q^2) ~ [ 2 ~ C_{n, {\rm LOG}}^{({\rm NLO})} + 
\Delta \gamma_{n, {\rm LOG}}^{(1, {\rm NS})} ] / 4 \pi$, so that 
Eq.~(\ref{eq:SGR}) reduces to the well-known NLO approximation. This 
implies that adopting the usual two-loop approximation for the running 
coupling constant $\alpha_s(Q^2)$, the twist-2 expression (\ref{eq:SGR}) 
contains all the NLO effects and the resummation of all the large-$n$ 
logarithms beyond the NLO.

The different running of the leading twist induced by resummation 
effects beyond the NLO has been investigated in Ref.~\cite{SIM00} for 
the unpolarized case, and in Ref.~\cite{simula_g1} for the moments of 
the proton $g_1$ structure function. It was found that, with respect 
to the NLO approximation, SGR effects enhance significantly the 
$Q^2$-evolution of the leading twist moments at $Q^2 \approx$~few~(GeV/c)$^2$, 
and that such an enhancement increases as the order $n$ of the moment increases.

As far as power corrections are concerned, several higher-twist operators 
exist and mix under the renormalization group equations. Such mixings are 
rather involved and the number of mixing operators increases with the order 
$n$ of the moment. A complete calculation of the higher-twist anomalous 
dimensions is not yet available, and therefore one has to use specific models 
or some phenomenological ansatz.

An interesting model for higher twists is the renormalon model 
\cite{renormalon}, which can be used as a guide to estimate the $x$-shape 
of the higher twists (or more precisely, of the twist-4 and twist-6 
terms). The renormalon model contains only one free-parameter, which means 
that it predicts the dependence of the higher-twist contribution to the 
moments upon the order $n$ up to an overall unknown constant. It is also 
characterized by the fact that the renormalon anomalous dimensions are the 
same as the leading twist ones. However, in Refs.~\cite{renormalon,Ricco1} 
it was already found that the renormalon model cannot explain simultaneously 
the power corrections to the transverse and longitudinal channels.
Moreover, several phenomenological extractions of higher-twist anomalous 
dimensions made in Refs.~\cite{Ji,Ricco1,SIM00,simula_g1,osipenko} 
suggest that the latter may differ significantly from the leading-twist 
ones. Therefore, in this work we use the same phenomenological ansatz as
adopted in Refs.~\cite{Ji,Ricco1,SIM00,simula_g1,osipenko} (and in
Ref.~\cite{M1p} for the $n=1$ moment), which does not exclude the 
renormalon picture, but is more general. 

To be specific, the Nachtmann moments are analyzed in terms of the
following twist expansion:
 \be
    M_n^N(Q^2) = \delta \eta_n(Q^2) + {\rm HT}_n(Q^2) ~ ,
    \label{eq:twists}
 \ee
where the higher-twist contribution ${\rm HT}_n(Q^2)$ is comprised of
twist-4 and twist-6 terms of the form
 \be
    \nonumber
    {\rm HT}_n(Q^2) &=& \delta a_n^{(4)}\biggl[\frac{\alpha_s(Q^2)}{\alpha_s(\mu^2)} 
    \biggr]^{\delta \gamma_n^{(4)}}\frac{\mu^2}{Q^2}\\
    &+& \delta a_n^{(6)}\biggl[\frac{\alpha_s(Q^2)}{\alpha_s(\mu^2)}\biggr]^{
    \delta \gamma_n^{(6)}}\frac{\mu^4}{Q^4} ~ ,
    \label{eq:HT}
 \ee
where the logarithmic pQCD evolution of the twist-$\kappa$ contribution 
is accounted for by the term $[\alpha_s(Q^2)]^{\delta \gamma_n^{(\kappa)}}$
with an {\em effective} anomalous dimension $\delta \gamma_n^{(\kappa)}$,
and the parameter $\delta a_n^{(\kappa)}$ represents the overall strength 
of the twist-$\kappa$ term at the renormalization scale $\mu^2$.

In Eq. (\ref{eq:HT}) only twist-4 and twist-6 terms are included.
In practice the number of higher-twist terms to be considered is mainly 
governed by the $Q^2$-range of the analysis. Indeed, as the latter is 
extended down to lower values of $Q^2$, more higher-twist terms are expected 
to contribute. Here we note that: ~ i) the inclusion of twist-4 and twist-6 
terms works well for $Q^2 \gtrsim 1$~(GeV/c)$^2$, as already found in the 
case of the unpolarized moments \cite{Ricco1,SIM00,osipenko}, and
~ ii) our least-$\chi^2$ fitting procedure turns out to be sensitive to
the presence of a twist-8 term only for $Q^2 \lesssim 1$~(GeV/c)$^2$,
where the resummation of high-order perturbative corrections may start 
to break down. Therefore, we limit ourselves to considering only twist-4 
and twist-6 terms in the analyses for $Q^2 \gtrsim 1$~(GeV/c)$^2$.

All the unknown parameters, namely the twist-2 coefficient $\delta A_n$,
as well as the four higher-twist parameters $\delta a_n^{(4)}, \delta 
\gamma_n^{(4)}, \delta a_n^{(6)}$ and $\delta \gamma_n^{(6)}$, are for each 
order $n$ simultaneously determined from a $\chi^2$-minimization procedure 
in the $Q^2$ range between $1$ and $30$~(GeV/c)$^2$. Changing the minimum 
$Q^2$ value down to $0.7 \div 0.8$~(GeV/c)$^2$ does not modify significantly 
the extracted values of the various twist parameters. On the other hand, 
increasing the minimum $Q^2$ up to $2$~(GeV/c)$^2$ leads to quite large 
uncertainties in the values of the twist parameters, due to a large 
decrease in the number of data points. 

The strong coupling constant in this analysis has been chosen to be
$\alpha_s(M_Z^2) = 0.118$, consistent with the twist analysis of the 
unpolarized moments made in Ref.~\cite{osipenko}. The (arbitrary) 
renormalization scale $\mu$ is set to $\mu = 1$~GeV/c. We point out 
that the high-$Q^2$ subset of the unpolarized Nachtmann moments of 
Ref.~\cite{osipenko} were analyzed in Ref.~\cite{alpha} in order to 
extract the value of $\alpha_s(M_Z^2)$, including SGR effects up to NLL 
accuracy. The value found, $\alpha_s(M_Z^2) = 0.1188 \pm 0.0010(\mbox{stat.}) 
\pm 0.0014(\mbox{sys.})$ (or $0.1188 \pm 0.0017$ adding the errors in 
quadrature), was in full agreement with the latest Particle Data Group 
world-average value $\alpha_s(M_Z^2) = 0.1187 \pm 0.0020$~\cite{PDG}.

The fitting procedure provides the best-fit values of the twist parameters 
together with their statistical uncertainties. The systematic uncertainties 
are, on the other hand, obtained by adding the systematic errors to the 
experimental moments and repeating the twist extraction procedure. Our 
results, including the uncertainties for each twist term separately, are 
reported in Table~\ref{table:twist1} and in Fig.~\ref{fig:twists}~\footnote{Note 
that for all the moments considered the data points at $Q^2 = 5$~(GeV/c)$^2$ 
are not reproduced by the twist expansion; in fact, their inclusion gives 
rise to extremely large values of $\chi^2$ for $n = 5$ and $n = 7$. The 
central values of the twist parameters reported in Table~\ref{table:twist1} 
are thus those obtained by excluding these data points in the fitting 
procedure, however, the impact of these points has been taken into account 
in the systematic errors in Table~\ref{table:twist1}.}. The ratio of the 
total higher-twist contribution, HT$_n(Q^2)$, to the leading twist term 
$\delta \eta_n(Q^2)$, is shown in Fig.~\ref{fig:ratio}(a). Note that since 
the leading twist component of the moments is directly extracted from the 
data, no specific functional shape for the leading twist parton distributions 
is assumed in our analysis. In the same way also our extracted higher twists 
do not rely upon any assumption about their $x$-shape.

\begin{table*}
\caption{\label{table:twist1}Leading twist $\delta \eta_n$ and 
higher-twist parameters, appearing in Eq.~(\ref{eq:HT}), extracted from 
the Nachtmann moments for $n \geq 3$ at the scale $Q^2 = 1$~(GeV/c)$^2$. 
The first errors are statistical, while the upper and lower ones are 
systematic.}
\begin{ruledtabular}
\begin{tabular}{|c|c|c|c|} \hline
                      & $M_3$                                   & $M_5$                                    & $M_7$                                    \\ \hline
$\delta \eta_n$       & $0.0147 \pm 0.0005_{-0.0023}^{+0.0025}$ & $0.0057 \pm 0.0008_{-0.0007}^{+0.0009}$  & $0.0038 \pm 0.0005_{-0.0002}^{+0.0003}$  \\ \hline
$\delta a_n^{(4)}$    & $0.020 \pm 0.001_{-0.007}^{+0.008}$     & $0.0155 \pm 0.0007_{-0.0009}^{+0.0047}$  & $0.0103 \pm 0.0005_{-0.0016}^{+0.0092}$  \\ \hline
$\delta \gamma^{(4)}$ & $2.2 \pm 0.3_{-0.9}^{+0.8}$             & $2.3 \pm 0.5_{-0.2}^{+0.5}$              & $2.6 \pm 0.4_{-0.1}^{+0.2}$              \\ \hline
$\delta a_n^{(6)}$    & $-0.012 \pm 0.002_{-0.007}^{+0.006}$    & $-0.0127 \pm 0.0009_{-0.0053}^{+0.0015}$ & $-0.0108 \pm 0.0005_{-0.0053}^{+0.0008}$ \\ \hline
$\delta \gamma^{(6)}$ & $3.0 \pm 0.6_{-1.5}^{+0.5}$             & $2.4 \pm 0.8_{-0.2}^{+0.1}$              & $2.9 \pm 0.5_{-0.2}^{+0.1}$              \\ \hline
\end{tabular}
\end{ruledtabular}
\end{table*}

\begin{figure*}
\includegraphics[bb=1cm 10cm 20cm 29cm, scale=0.75]{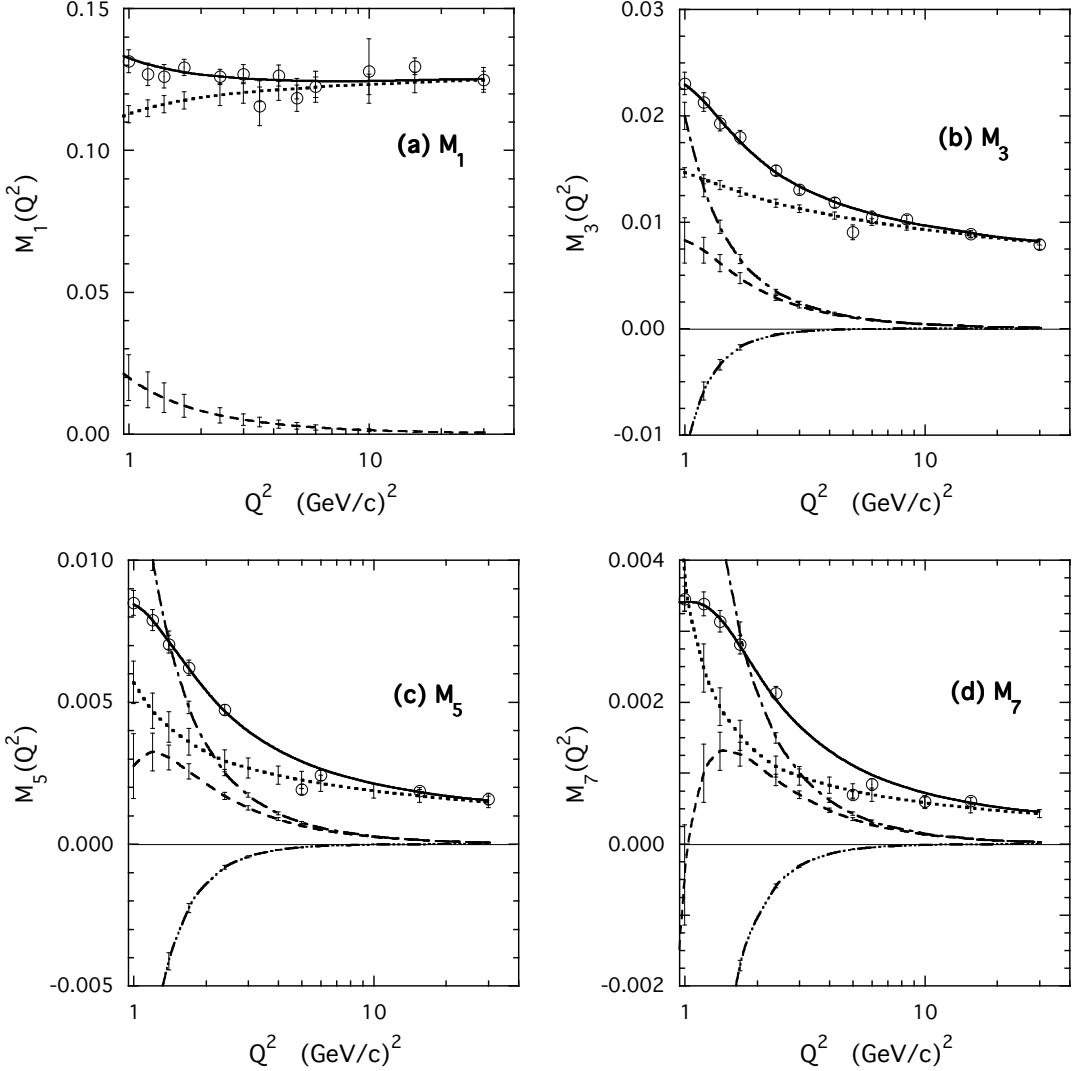}
\caption{\label{fig:twists} Results of the twist analysis for $n = 1$ 
(adapted from Ref.~\cite{M1p}) and for $n = 3, 5$ and $7$ obtained in 
this work. Open circles represent the Nachtmann moments, and the solid 
lines are fits to the moments using Eqs.~(\ref{eq:twists}), (\ref{eq:HT}) 
and (\ref{eq:SGR}) with the parameters listed in Table~\ref{table:twist1}. 
The twist-2 (dotted), twist-4 (dot-dashed), twist-6 (triple-dot-dashed) 
and total higher twist (dashed) contributions are shown separately. The 
errors indicated are statistical.}
\end{figure*}

\begin{figure}
\includegraphics[bb=0.5cm 4.5cm 20cm 29cm, scale=0.65]{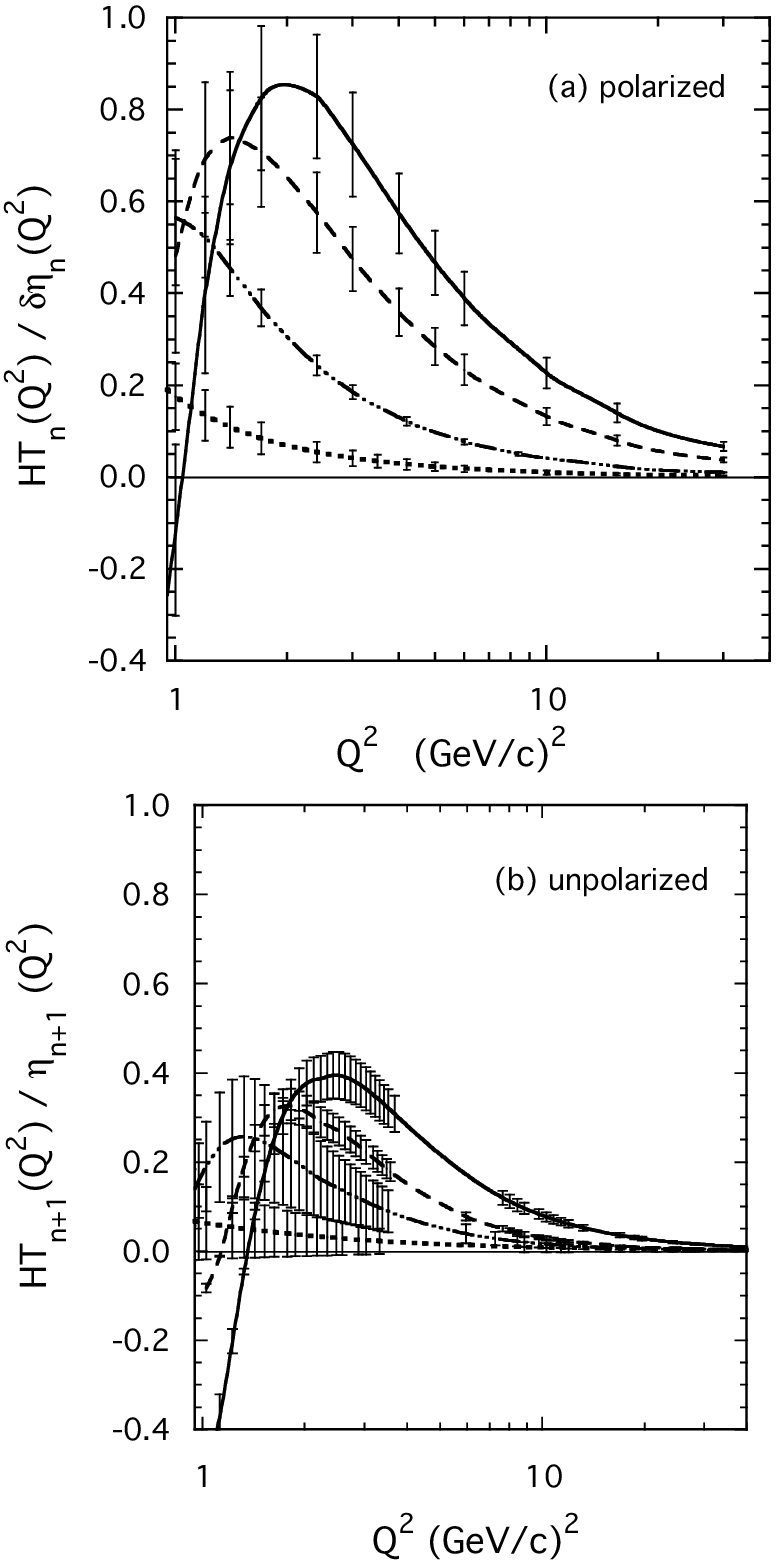}
\caption{\label{fig:ratio}(a) Ratio of the total higher-twist [see 
Eq.~(\ref{eq:HT})] to the leading twist given in Eq.~(\ref{eq:SGR}). 
Dotted line - $M_1$ (from Ref.~\cite{M1p}); triple-dot-dashed line - $M_3$;
dashed line - $M_5$; solid line - $M_7$. (b) Ratio of the total higher 
twist to the leading twist obtained in the analysis of the unpolarized 
moments in Ref.~\cite{osipenko}.}
\end{figure}

Our main results for the higher twists in
Figs.~\ref{fig:twists}--\ref{fig:ratio} can be summarized as follows:

\begin{itemize}

\item The extracted twist-2 term yields an important contribution in
the whole $Q^2$-range of the present analysis; it is determined quite 
accurately with an uncertainty which does not exceed 15\% (statistical) 
and 20\% (systematic);

\item The $Q^2$-dependence of the data leaves room for a higher-twist
contribution which runs slower than a pure $1 / Q^2$ dependence, or may
even become negative at the lowest values of $Q^2$ and large $n$.
This requires in Eq.~(\ref{eq:HT}) a twist-6 term with a sign opposite 
to that of the twist-4. As already noted in 
Refs.~\cite{Ricco1,SIM00,simula_g1}, such opposite signs make the 
total higher-twist contribution smaller than its individual terms
(see dashed lines in Fig.~\ref{fig:twists});

\item The extracted values of the higher-twist anomalous dimensions
appear to be significantly larger than the corresponding ones of the 
leading twist (viz., $\gamma_n^{\rm NS} = 0.67, 0.97, 1.17$ for
$n = 3, 5, 7$, respectively, at $N_f = 4$);

\item The total higher-twist contribution is important for
$Q^2 \approx$~few~(GeV/c)$^2$, and is still non-negligible even at
$Q^2 \simeq 10$~(GeV/c)$^2$ for the higher moments.
Comparison with the higher twists extracted from the moments of the
unpolarized $F_2$ structure function \cite{osipenko} in
Fig.~\ref{fig:ratio} clearly shows that the total higher-twist 
contribution is significantly larger in the polarized case, 
as already observed in Ref.~\cite{simula_g1} and also in agreement
with the findings of Ref.~\cite{Leader}.

\end{itemize}

The extracted twist-2 contribution is given in Table~\ref{table:ltw}
and in Fig.~\ref{fig:twist-2}, where it is compared with several NLO
parameterizations of spin-dependent parton distribution functions (PDFs)
\cite{GRSV,DS,BB,LSS}. For $n = 1$ the twist-2 moment obtained in 
Ref.~\cite{M1p} agrees well at large $Q^2$ with the results of 
Refs.~\cite{BB,LSS}, whereas at lower $Q^2$ our findings are below 
the predictions of all the four PDF sets. We should note, however, that 
in Ref.~\cite{M1p} a next-to-next-to-next-to-leading order (N$^3$LO) 
approximation was adopted, since for the $n = 1$ moment the SGR effects 
are totally absent. This gives rise to a running of the leading twist 
which is faster than that at NLO. As $n$ increases, our extracted twist-2 
runs faster around $Q^2 \approx$~few~(GeV/c)$^2$, in agreement with the 
findings of Refs.~\cite{simula_g1,SIM00},i.e.~the running is enhanced by 
SGR effects with respect to the NLO scheme adopted in 
Refs.~\cite{GRSV,DS,BB,LSS}.

Note that at large $Q^2$ ($\gtrsim 10$~(GeV/c)$^2$) the extracted
twist-2 contributions for $n > 1$ in Fig.~\ref{fig:twist-2} is
systematically below the parameterizations in Refs.~\cite{GRSV,DS,BB,LSS},
with the discrepancy increasing with the order $n$. This would imply 
PDFs lower than those of Refs.~\cite{GRSV,DS,BB,LSS} at large $x$.
Such an effect may at least partially be due to the neglect, or a
different treatment, of higher-twist effects in the analyses of
Refs.~\cite{GRSV,DS,BB,LSS}, which were carried out in $x$-space
(see {\em e.g.}, Ref.~\cite{Leader}). To fully unravel the origin of 
the above differences is, however, beyond the aim of the present paper.

\begin{table*}
\caption{\label{table:ltw} The extracted leading twist contribution 
  $\eta_n(Q^2)$ [see Eq.~(\ref{eq:SGR})], reported with statistical and 
  systematic errors.}
\begin{ruledtabular}
\begin{tabular}{|c|c|c|c|c|} \cline{1-5}
$Q^2~[$(GeV/c)$^2]$ & $\delta \eta_1(Q^2)$ & $\delta \eta_3(Q^2) \times 10^{-2}$ & $\delta \eta_5(Q^2) \times 10^{-2}$ & $\delta \eta_7(Q^2) \times 10^{-2}$ \\ \cline{1-5}
 1.00 &  $0.1127 \pm 0.0030 \pm 0.0109$  &  $1.47 \pm 0.05 \pm 0.24$  &  $0.57 \pm 0.08 \pm 0.08$  &  $0.380 \pm 0.052 \pm 0.048$  \\ \cline{1-5}
 1.20 &  $0.1148 \pm 0.0030 \pm 0.0109$  &  $1.40 \pm 0.04 \pm 0.23$  &  $0.47 \pm 0.06 \pm 0.06$  &  $0.248 \pm 0.034 \pm 0.031$  \\ \cline{1-5}
 1.40 &  $0.1162 \pm 0.0030 \pm 0.0108$  &  $1.35 \pm 0.04 \pm 0.22$  &  $0.41 \pm 0.05 \pm 0.05$  &  $0.193 \pm 0.026 \pm 0.024$  \\ \cline{1-5}
 1.70 &  $0.1176 \pm 0.0030 \pm 0.0108$  &  $1.28 \pm 0.04 \pm 0.21$  &  $0.36 \pm 0.05 \pm 0.05$  &  $0.153 \pm 0.021 \pm 0.019$  \\ \cline{1-5}
 2.40 &  $0.1195 \pm 0.0037 \pm 0.0135$  &  $1.18 \pm 0.04 \pm 0.19$  &  $0.29 \pm 0.04 \pm 0.04$  &  $0.109 \pm 0.015 \pm 0.014$  \\ \cline{1-5}
 3.00 &  $0.1203 \pm 0.0037 \pm 0.0134$  &  $1.13 \pm 0.03 \pm 0.18$  &  $0.27 \pm 0.04 \pm 0.04$  &  $0.096 \pm 0.013 \pm 0.012$  \\ \cline{1-5}
 3.50 &  $0.1208 \pm 0.0037 \pm 0.0134$  &  $1.10 \pm 0.03 \pm 0.18$  &  $0.25 \pm 0.03 \pm 0.03$  &  $0.088 \pm 0.012 \pm 0.011$  \\ \cline{1-5}
 4.20 &  $0.1213 \pm 0.0037 \pm 0.0133$  &  $1.06 \pm 0.03 \pm 0.17$  &  $0.24 \pm 0.03 \pm 0.03$  &  $0.081 \pm 0.011 \pm 0.010$  \\ \cline{1-5}
 5.00 &  $0.1217 \pm 0.0037 \pm 0.0133$  &  $1.03 \pm 0.03 \pm 0.17$  &  $0.23 \pm 0.03 \pm 0.03$  &  $0.075 \pm 0.010 \pm 0.009$  \\ \cline{1-5}
 6.00 &  $0.1222 \pm 0.0036 \pm 0.0133$  &  $1.00 \pm 0.03 \pm 0.16$  &  $0.21 \pm 0.03 \pm 0.03$  &  $0.070 \pm 0.010 \pm 0.009$  \\ \cline{1-5}
 8.40 &  $0.1229 \pm 0.0036 \pm 0.0132$  &  $0.95 \pm 0.03 \pm 0.16$  &  $0.20 \pm 0.03 \pm 0.03$  &  $0.062 \pm 0.008 \pm 0.008$  \\ \cline{1-5}
10.00 &  $0.1232 \pm 0.0036 \pm 0.0132$  &  $0.93 \pm 0.03 \pm 0.15$  &  $0.19 \pm 0.02 \pm 0.02$  &  $0.058 \pm 0.008 \pm 0.007$  \\ \cline{1-5}
15.50 &  $0.1239 \pm 0.0036 \pm 0.0132$  &  $0.88 \pm 0.03 \pm 0.14$  &  $0.17 \pm 0.02 \pm 0.02$  &  $0.051 \pm 0.007 \pm 0.006$  \\ \cline{1-5}
30.00 &  $0.1247 \pm 0.0032 \pm 0.0115$  &  $0.81 \pm 0.03 \pm 0.13$  &  $0.15 \pm 0.02 \pm 0.02$  &  $0.043 \pm 0.006 \pm 0.005$  \\ \cline{1-5}
\end{tabular}
\end{ruledtabular}
\end{table*}

\begin{figure*}
\includegraphics[bb=1cm 10cm 20cm 29cm, scale=0.75]{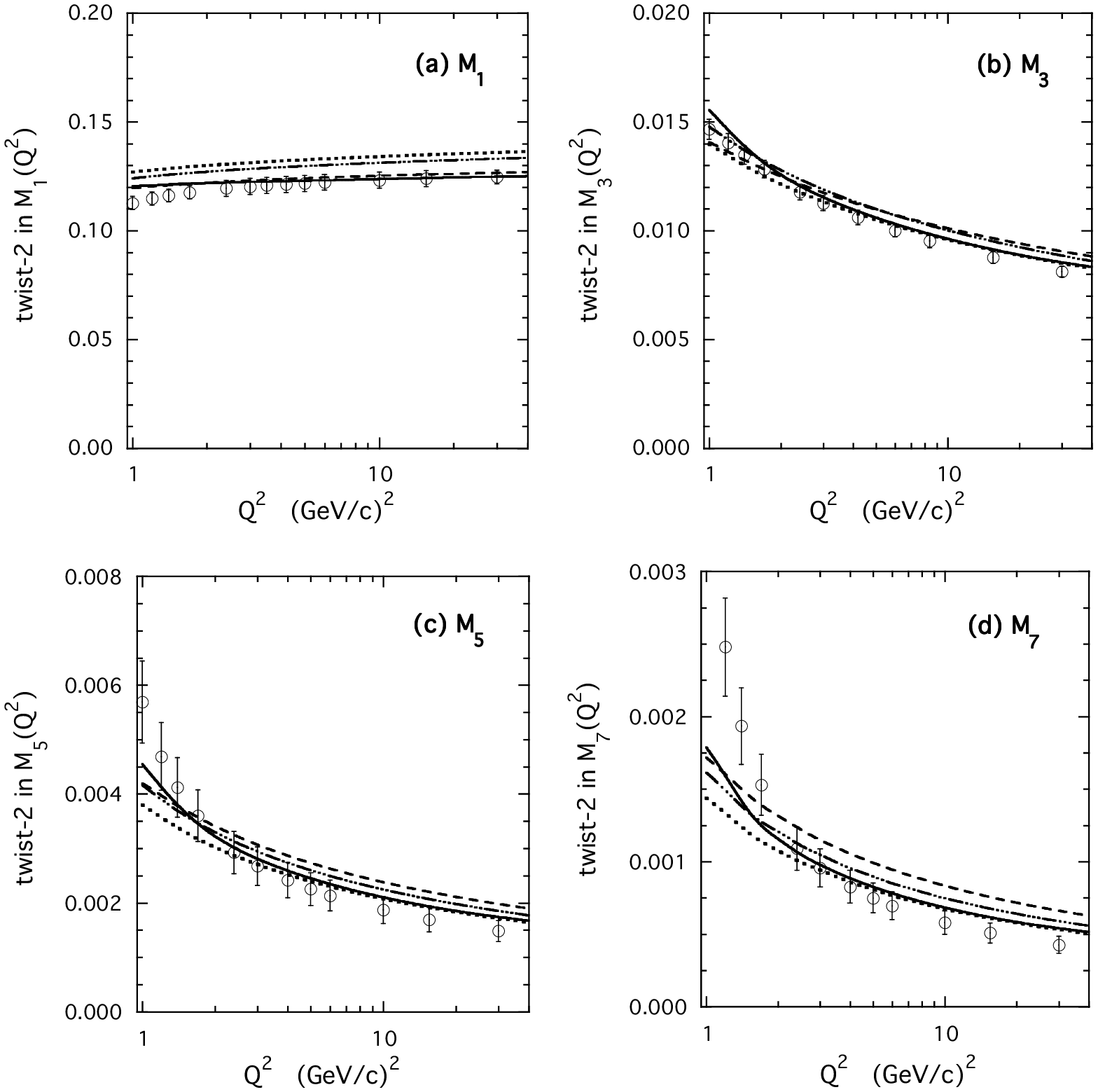}
\caption{\label{fig:twist-2}
  The leading twist moments (open circles) extracted in the present
  analysis for $n \geq 3$ and in Ref.~\cite{M1p} for $n = 1$, compared 
  with the corresponding moments of various parton distribution sets:
  dotted~\cite{GRSV};
  triple-dot-dashed~\cite{DS};
  dashed~\cite{BB};
  solid~\cite{LSS}.}
\end{figure*}

\section{Conclusions}\label{sec:Conclusions}

We have presented a self-consistent analysis of world data on the proton 
$g_1$ structure function in the range $0.17 < Q^2 < 30$~(GeV/c)$^2$, 
including recent measurements performed with the CLAS detector at 
Jefferson Lab \cite{fatemi}. This analysis has made it possible to 
accurately compute for the first time the low-order moments of $g_1$ 
and study their evolution from small to large values of $Q^2$. Our analysis 
includes the latest experimental results from Jefferson Lab for the ratio 
$R = \sigma_L / \sigma_T$ and a new model for the transverse asymmetry 
$A_2$ in the resonance production regions, as well as the unpolarized cross 
sections measured recently in the resonance region at Jefferson 
Lab~\cite{osipenko,hallc}.

Within the framework of the operator product expansion, we have extracted
from the experimental moments at $Q^2 \gtrsim 1$~(GeV/c)$^2$ the
contributions of both leading and higher twists. Effects from radiative 
corrections beyond the next-to-leading order have been taken into 
account by means of soft-gluon resummation techniques.

The leading twist has been determined with good accuracy, allowing detailed 
comparisons to be made with various NLO polarized parton distribution functions 
obtained from global analyses in Bjorken-$x$ space. A faster running in $Q^2$ 
is observed in our twist-2 moments due to the inclusion of resummation effects 
beyond NLO. The twist-2 moments are also found to lie slightly below those 
calculated from the standard polarized PDFs, suggesting that the latter 
overestimate the leading twist at large $x$. This may reflect the different 
treatment of higher-twist effects in our analysis compared with those in the 
global PDF fits.

The contribution of higher twists to the polarized proton structure
function $g_1$ is found to be significantly larger than for the unpolarized 
proton structure function $F_2$, although some cancellations between different 
twists occurs at low $Q^2$.

Improvements in the determination of both the leading and higher twist
terms are expected to come with the availability of new CLAS data taken 
at Jefferson Lab with the 6~GeV electron beam, which will provide an 
extended kinematical coverage up to $Q^2 \approx 5$~(GeV/c)$^2$. Beyond 
this, we anticipate significant progress in the measurement of polarized 
structure functions at higher $Q^2$ and over a larger range of $x$ with 
the upgrade of the Jefferson Lab electron beam to 12~GeV.

\begin{acknowledgments}
This work was supported by the Istituto Nazionale di Fisica Nucleare, the French 
Commissariat  \`a l'Energie Atomique, French Centre National de la Recherche Scientifique, 
the U.S.~Department of Energy and National Science Foundation and the Korea Science and 
Engineering Foundation. The Southeastern Universities Research Association (SURA) operates 
the Thomas Jefferson National Accelerator Facility for the United States Department of 
Energy under contract DE-AC05-84ER40150.

\end{acknowledgments}

\appendix

\section{Fit of the proton transverse asymmetry $A_2$}\label{appendix1}

The parameterization of $A_2$ is based on an estimate of the polarized 
transverse structure function $g_T$ by means of resonance-background 
separation, where the resonance part is taken from a constituent quark 
(CQ) model~\cite{Giannini}, while the background is described according 
Wandzura-Wilczek (WW) prescription \cite{WW}. As normalization, we use 
the Burkhardt-Cottingham (BC) sum rule \cite{BC}, for each $Q^2$ value of 
the data. The BC sum rule implies that
\begin{equation}
\label{eq:BC}
\int_0^1 dx\ g_2(x,Q^2) = 0
\end{equation}
for any $Q^2$, where the integration includes also the elastic peak.

In practice it is more convenient to work with the purely transverse
structure function $g_T$, which is defined as
 \be
    g_T(x,Q^2) = g_1(x,Q^2) + g_2(x,Q^2)
 \ee
Decomposing $g_T$ into leading twist, elastic and higher twist terms,
we can write
 \be
    \label{eq:sep}
    g_T(x,Q^2) & = & g_T^{\rm WW}(x,Q^2)
    +  g_T^{\rm el}(Q^2) ~ \delta (1-x) \nonumber \\
    & + & ~ g_T^{\rm HT}(x,Q^2)
 \ee
where the first term represents the (twist-2) WW relation (which is found
to be a good approximation in DIS), the second term represents the elastic 
peak contribution, and the third parameterizes the remaining (higher twist)
part of $g_T$. 

Next we make use of an {\em ansatz} which assumes that the first term in 
Eq.~(\ref{eq:sep}), $g_T^{\rm WW}(x,Q^2)$, is due to the background 
contribution and the second term, $g_T^{\rm HT}(x,Q^2)$, contains only 
the resonance part of the total cross section,
 \be
    g_T^{\rm WW}(x,Q^2) & = & g_T^{\rm bkg}(x,Q^2)\ , \\
    g_T^{\rm HT}(x,Q^2) & = & g_T^{\rm res}(x,Q^2)\ .
 \ee
This {\em ansatz} is motivated partly by duality arguments \cite{DUAL_OLD} 
as well as by recent findings in polarized structure function studies, which 
suggest a picture in which the resonance peaks fluctuate around a smooth 
background extrapolated from the DIS regime. Clearly this model neglects 
the interference between resonances and the background, which can play 
an important role in the total cross section. However, given the absence 
of experimental guidance (at least above the two-pion production threshold), 
this approach is the minimal one suitable for the present analysis.

Using the WW relation \cite{WW}, one can rewrite $g_T$ in
Eq.~(\ref{eq:sep}) as
 \be
    g_T(x,Q^2) & = & \int_x^{x_{\rm th}} \frac{dy}{y} g_1(y,Q^2)
    \nonumber \\
    & + & g_T^{\rm el}(Q^2)\ \delta (1-x)\ +\ g_T^{\rm HT}(x,Q^2)\ .
\ee
\noindent From the BC sum rule in Eq.~(\ref{eq:BC}) and the Fubini
theorem \cite{Fubini} we then find:
 \be
    && \int_x^{x_{\rm th}} dx/ g_T^{\rm HT}(x,Q^2) = g_1^{\rm el}(Q^2) - 
    g_T^{\rm el}(Q^2) \nonumber\\
    && = \frac{Q^2}{8M^2+2Q^2}G_M(Q^2) (G_M(Q^2)-G_E(Q^2)) ~~~~ 
    \label{eq:BC_gt}
 \ee
where $G_E(Q^2)$ and $G_M(Q^2)$ are the Sachs proton electric and magnetic 
form factors.

The WW term $g_T^{\rm WW}$ is calculated from the phenomenological
parameterization of $g_1$ given in Ref.~\cite{simula_g1}, which is known 
to work well also in the resonance region and at the photon point ($Q^2 = 0$).
Furthermore, target mass corrections are applied in order to remove the 
kinematical effects of working at finite $Q^2$,
 \be
    g_T^{\rm WW-TMC} & = & \frac{1}{r^2}\frac{x}{\xi} \int_{\xi}^{\xi_{\rm th}} 
    d\xi^\prime\ \frac{g_1(\xi^\prime)}{\xi^\prime} 
    \nonumber \\
    & + & \frac{2M^2}{Q^2}\frac{x^2}{r^3} \int_{\xi}^{\xi_{\rm th}} 
    d\xi^\prime\ \frac{g_1(\xi^\prime)}{\xi^\prime} 
    \log{\frac{\xi^\prime}{\xi}}\ , ~~~~
 \ee
where $r = \sqrt{1 + 4 M^2 x^2 / Q^2}$. The resonance part of $g_T$ is directly 
related to the longitudinal-transverse interference term of the resonance 
production cross section,
 \be
    g_T^{\rm res} (W,Q^2) = - \frac{\nu M K}{4 \pi^2 \alpha \sqrt{Q^2}} 
    \sigma^{LT^\prime} (W,Q^2)
    \label{eq:gt_res}
 \ee
where
 \be
    \sigma^{LT^\prime}(W,Q^2) & = & \sum_{N^*} \pi \frac{M \sqrt{2 Q^2}}{W q^*} 
    B(W) \nonumber \\
    & \times & S^*_{1/2}(Q^2) A_{1/2}(Q^2)\ .
 \ee
Here the sum runs over all nucleon excited states $N^*$, $B(W)$ is the 
unit-area resonance shape described in the relativistic Breit-Wigner 
approximation,
 \be
    B(W) = \frac{W M_{\rm res}}{\pi} \frac{\Gamma_{\rm res}}{(W^2 - 
    M_{\rm res}^2)^2+M_{\rm res}^2 \Gamma_{\rm res}^2}\ ,
 \ee
and $q^*$ is the 3-momentum transfer in the resonance rest frame,
 \be
    q^* = \Biggl\{ Q^2+ \frac{W^2-M^2-Q^2}{4 W^2} \Biggr\}^{1/2}\ .
 \ee

The helicity amplitude $A_{1/2}(Q^2)$ is relatively well known for the
most prominent resonances, while the longitudinal amplitude $S_{1/2}(Q^2)$
is largely unexplored experimentally, apart from the $\Delta(1232)$
resonance for which some data do exist. Theoretical predictions for these 
amplitudes can be obtained from CQ models which successfully describe 
resonance mass spectra and some transverse electromagnetic couplings.
We use the CQ model from Ref.~\cite{Giannini} for both the $A_{1/2}(Q^2)$
and $S_{1/2}(Q^2)$ amplitudes in order to calculate $g_T^{\rm res}$ in 
Eq.~(\ref{eq:gt_res}).

Unfortunately, the $Q^2$-evolution of the couplings $A_{1/2}(Q^2)$ and 
$S_{1/2}(Q^2)$ in CQ models depends strongly on the choice of the
potential and other model parameters. In order to improve this description 
we apply the BC sum rule given in Eqs.~(\ref{eq:BC}) and (\ref{eq:BC_gt}) 
to the entire resonance part of  $g_T^{\rm res}$. This amounts to modifying 
$g_T^{\rm res}$ by multiplying it by a factor
 \be
     N(Q^2) = \frac{g_1^{\rm el}(Q^2) - g_T^{\rm el}(Q^2)}
     {\int_0^{x_{\rm th}} dx\ g_T^{\rm res}(x,Q^2)}\ .
    \label{eq:factor}
 \ee
Therefore, at each given $Q^2$ the BC sum rule defines the total area
of the resonance structure function $g_T^{\rm res}$.

The asymmetry $A_2$ can then be directly related to $g_T$ according to
 \be
    A_2(x,Q^2) = \frac{\sqrt{Q^2}}{\nu} \frac{g_T(x,Q^2)}{F_1(x,Q^2)} ~~,
 \ee
where $F_1(x,Q^2)$ is the familiar unpolarized structure function. The 
final parameterization is shown in Fig.~\ref{fig:f1}, compared with
calculations of the MAID model from Ref.~\cite{MAID}. The MAID results 
represent a sum over a few exclusive channels which should be reliable 
when $W$ is not very large. New experimental data on $g_2$ in the resonance 
region at different $Q^2$ values are clearly needed.

\begin{figure}[htb]
\begin{center}
\includegraphics[bb=0.25cm 5.5cm 20cm 23cm, scale=0.45]{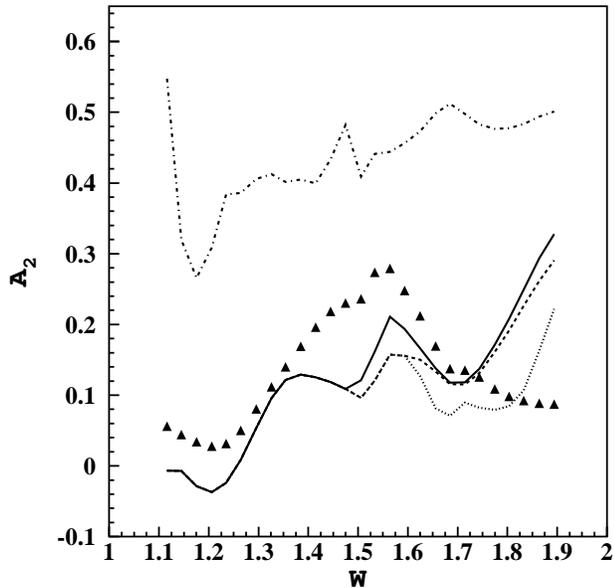}
\caption{\label{fig:f1} Constituent quark model calculations of $A_2(W, Q^2)$ 
in comparison with the MAID model predictions~\cite{MAID} at $Q^2 = 
1.3$~(GeV/c)$^2$: triangles show the calculations as described in the text; 
solid ($\pi$ production), dashed ($\pi$ and $\eta$) and dotted ($\pi$, $\eta$, 
$K\Lambda$ and $K\Sigma$) lines represent MAID model calculations. The 
dot-dashed curve indicates the upper Soffer limit on $A_2$.}
\end{center}
\end{figure}

In the DIS region data from 
Refs.~\cite{SLAC-E143,SLAC-E155,SLAC-E155x,SMC-NA47b} suggest that $A_2$ 
is rather small, and can be described within the WW approach. In order to 
quantify the agreement and to estimate the systematic uncertainty, we plot 
in Fig.~\ref{fig:f3} the weighted difference between the data and the WW 
prescription,
 \be
    \Delta A_2 = \frac{A_2^{\rm exp} - A_2^{\rm WW}}{\delta_{A_2}^2}\ ,
 \ee
where $\delta_{A_2}$ is the $A_2^{\rm exp}$ statistical error. One sees 
that the mean value within errors is compatible with zero, and the error 
of $4 \times 10^{-2}$ has been estimated according the formula
 \be
    \delta^{\rm sys} (A_2) = \Biggl[\sum_i^N 
    \frac{1}{\delta_{A_2}^2(x_i,Q^2_i)}\Biggr]^{-1/2} \sigma_{\Delta A_2}\ ,
 \ee
where $\sigma_{\Delta A_2}$ is the width of the $\Delta A_2$ distribution 
and the sum runs over all available $A_2$ experimental points ($N$). Therefore, 
in the DIS kinematics, defined here as $W > 2$~GeV, the asymmetry $A_2$ can 
be estimated through the WW formula within the systematic uncertainty of 
$\delta^{\rm sys} (A_2) = 4 \times 10^{-2}$. However, taking into account 
target mass corrections, which affect the $g_T$ structure function also 
in the DIS region, one finally finds $\delta^{\rm sys} (A_2) = 1.6 \times 
10^{-2}$ [see Fig.~\ref{fig:f3}].

\begin{figure}[thb]
\begin{center}
\includegraphics[bb=0.25cm 5.5cm 20cm 23cm, scale=0.45]{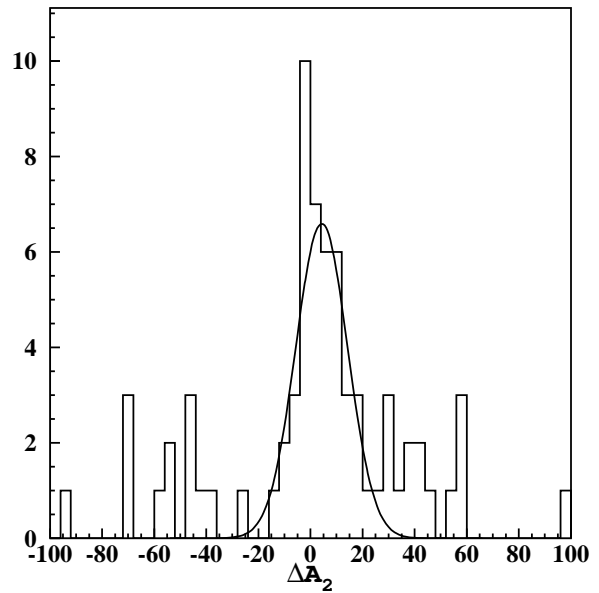}
\caption{\label{fig:f3} Weighted difference between the experimental $A_2$ 
values and the WW prescription $A_2^{\rm WW}$ including the target mass 
corrections.}
\end{center}
\end{figure}

\section{KINEMATIC HIGHER TWISTS}\label{appendix2}

In order to estimate contribution of the kinematic twists appearing in 
the expansion of the CN moments, we extract from our data the inelastic 
part of the $d_2$ moment, defined as
 \be
    d_2(Q^2) = \int_0^1 dx\ x^2 \Bigl\{ 3 g_T(x,Q^2) - g_1(x,Q^2) \Bigr\}\ ,
 \ee
where the structure function $g_T(x,Q^2)$ is described in 
Appendix~\ref{appendix1}. The extracted values of $d_2(Q^2)$ are given in 
Table~\ref{table:dn} and shown in Fig.~\ref{fig:d2}. 

\begin{table}[htb]
\caption{\label{table:dn}The inelastic part of $d_2(Q^2)$ extracted from 
data (see text). The results are reported together with the statistical 
and systematic errors.}

\vspace{0.25cm}

\begin{tabular}{|c|c|} \cline{1-2}
$Q^2~ [$GeV$^2]$ & $d_2(Q^2) \cdot 10^{-3}$ \\ \cline{1-2}
 0.17 &  3.7  $\pm$ 1.6 $\pm$  2.1 \\ \cline{1-2}
 0.20 &  3.9  $\pm$ 0.8 $\pm$  2.5 \\ \cline{1-2}
 0.24 &  4.9  $\pm$ 1   $\pm$  3.6 \\ \cline{1-2}
 0.30 &  8    $\pm$ 1   $\pm$  4.8 \\ \cline{1-2}
 0.35 &  9.3  $\pm$ 0.9 $\pm$  5.4 \\ \cline{1-2}
 0.42 & 10.2  $\pm$ 2   $\pm$  6.3 \\ \cline{1-2}
 0.50 & 12.3  $\pm$ 1.8 $\pm$  8   \\ \cline{1-2}
 0.60 & 14.4  $\pm$ 1.4 $\pm$  9   \\ \cline{1-2}
 0.70 & 14.6  $\pm$ 1.2 $\pm$  9.6 \\ \cline{1-2}
 0.84 & 14.4  $\pm$ 1.2 $\pm$ 10   \\ \cline{1-2}
 1.00 & 14.4  $\pm$ 1   $\pm$ 11   \\ \cline{1-2}
 1.20 & 11.6  $\pm$ 1.2 $\pm$ 11   \\ \cline{1-2}
 1.40 & 10    $\pm$ 1.2 $\pm$ 11   \\ \cline{1-2}
 1.70 &  6.8  $\pm$ 1.5 $\pm$ 11   \\ \cline{1-2}
 2.40 &  3.7  $\pm$ 1.3 $\pm$ 12   \\ \cline{1-2}
 3.00 &  2.9  $\pm$ 1   $\pm$ 12   \\ \cline{1-2}
 3.50 &  3.9  $\pm$ 0.5 $\pm$ 17   \\ \cline{1-2}
 4.20 &  1.4  $\pm$ 1.1 $\pm$ 13   \\ \cline{1-2}
 5.00 &  3.5  $\pm$ 1.6 $\pm$ 15   \\ \cline{1-2}
 6.00 &  1.3  $\pm$ 1.3 $\pm$ 15   \\ \cline{1-2}
10.00 &  1.7  $\pm$ 1.2 $\pm$ 18   \\ \cline{1-2}
15.50 &  0.6  $\pm$ 0.7 $\pm$ 22   \\ \cline{1-2}
30.00 &  0.3  $\pm$ 0.9 $\pm$ 30   \\ \cline{1-2}
\end{tabular}

\vspace{0.25cm}

\end{table}

\begin{figure}[htb]
\begin{center}
\includegraphics[bb=0.25cm 5.5cm 20cm 23cm, scale=0.45]{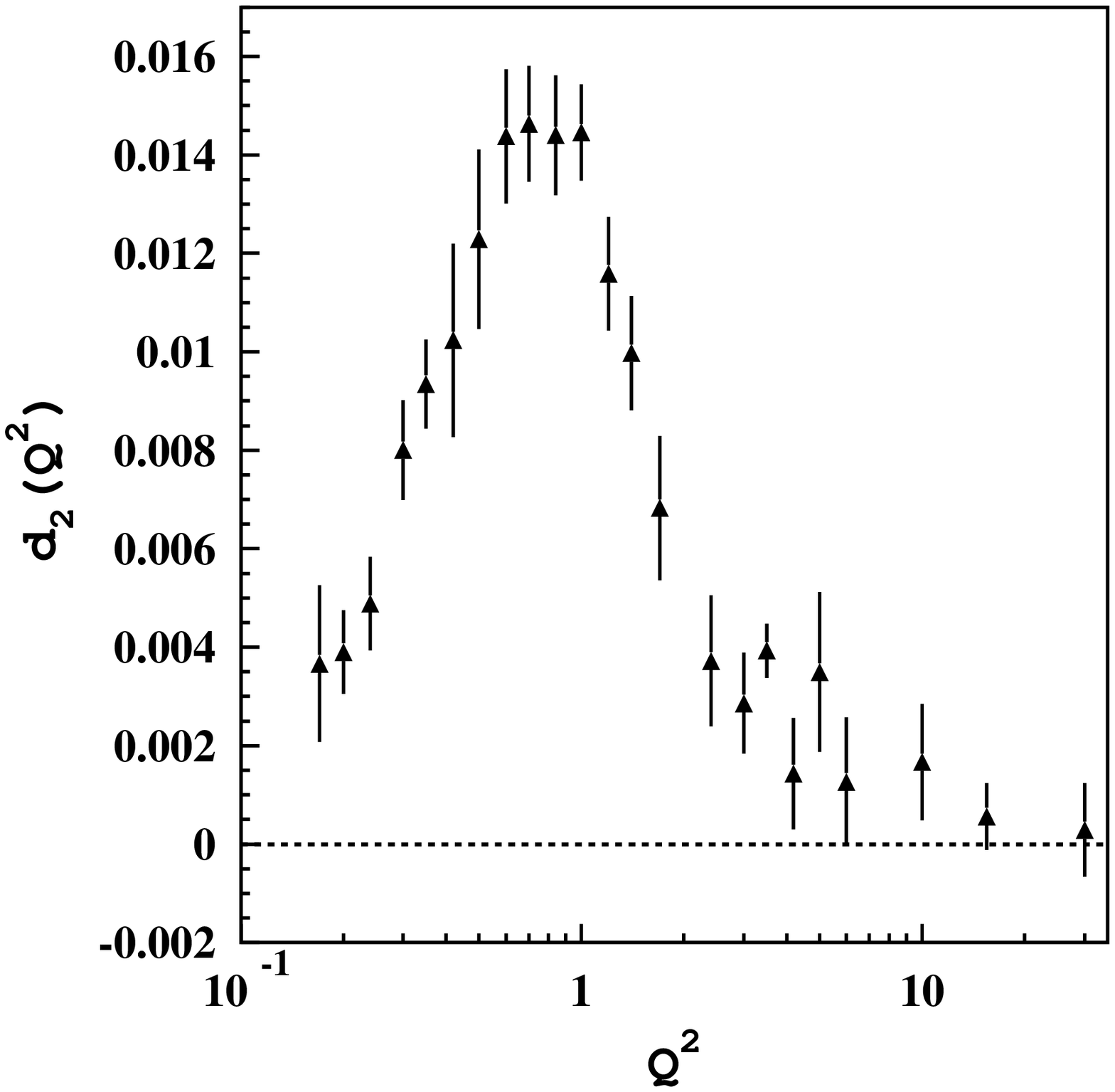}
\caption{\label{fig:d2} Extracted $d_2$ contribution to the first moment 
of the structure function $g_1$.}
\end{center}
\end{figure}

The lowest twist component in $d_2$ is twist-3, although higher twists can 
also contribute to $d_2$ at low $Q^2$. Note that only the inelastic part of 
$d_2$ is extracted; the elastic contribution has 
to be added separately for a twist analysis of $d_2$. The results indicate that 
at high $Q^2$ the values of $d_2(Q^2)$ are consistent with a vanishing 
twist-3 contribution.

\end{document}